\documentclass[]{pasj01}
\draft

\usepackage{multirow}

\begin{document} 
\Received{}
\Accepted{}

\title{The mass--metallicity relation AKARI-FMOS infrared galaxies at $z\sim0.88$ in the AKARI North Ecliptic Pole Deep Survey Field}

\author{Nagisa \textsc{Oi}\altaffilmark{1, 2, 8}%
}
\email{nagisaoi@rs.tus.ac.jp, nagisaoi@ir.isas.jaxa.jp}

\author{Tomotsugu \textsc{Goto}\altaffilmark{3}}

\author{Matthew \textsc{Malkan}\altaffilmark{4}}

\author{Chris \textsc{Pearson}\altaffilmark{5, 6, 7}}

\author{Hideo \textsc{Matsuhara}\altaffilmark{8}}

\altaffiltext{1}{Kwansei Gakuin University, 2-1, Gakuen, Sanda, Hyogo 669-1337 Japan}
\altaffiltext{2}{Tokyo University of Science, 1-3, Kagurazaka, Shinjuku-ku, Tokyo, 162-8601 Japan}
\altaffiltext{3}{National Tsing hua University, No. 101, Section 2, Kuang-Fu Road, Hsinchu, Taiwan 30013}
\altaffiltext{4}{Department of Physics and Astronomy, UCLA, Los Angeles, CA, 90095-1547, USA}
\altaffiltext{5}{RAL Space, CCLRC Rutherford Appleton Laboratory, Chilton, Didcot, Oxfordshire OX11 0QX, UK}
\altaffiltext{6}{Department of Physical Sciences, The Open University, Milton Keynes, MK7 6AA, UK}
\altaffiltext{7}{Oxford Astrophysics, Denys Wilkinson Building, University of Oxford, Keble Rd, Oxford OX1 3RH, UK}
\altaffiltext{8}{Department of Infrared Astrophysics,
         Institute of Space and Astronautical Science,
         Japan Aerospace Exploration Agency, 3-1-1 Yoshinodai, Chuo-ku, Sagamihara 252-5210 Japan }


\maketitle

 \begin{abstract}
  Mass, metallicity, and star formation rate (SFR) of a galaxy are crucial parameters in understanding galaxy formation and evolution. However, the relation among these, (i.e., the fundamental relation) is still a matter of debate for luminous infrared galaxies, which carry a bulk of star formation rate budget of the universe at z$\sim$1.
We have investigated the relation among stellar mass, gas-phase oxygen abundance, and SFR  of AKARI-detected mid-IR galaxies at z$\sim$ 0.88 in the AKARI NEP deep field.
  We observed $\sim$350 AKARI sources with Subaru/FMOS near-infrared spectrograph, and detected sure H$\alpha$ emission lines from 25 galaxies and expected H$\alpha$ emission lines from 44 galaxies.
  The $SFR_{H\alpha,IR}$ of our sample is almost constant ($\langle~SFR_{H\alpha,IR}~\rangle$~=~$\sim25$~${\rm M}_{\Sol}~{\rm yr}^{-1}$) over the stellar mass range of our sample.
Compared with main-sequence (MS) galaxies at a similar redshift range ($z\sim0.78$), the average SFR of our detected sample is comparable for massive galaxies ($\sim$10$^{10.58}~{\rm M}_{\Sol}$), while higher by $\sim$0.6~dex for less massive galaxies ($\sim$10$^{10.05}~{\rm M}_{\Sol}$).
We measure metallicities from the [N \emissiontype{II}]/H$\alpha$ emission line ratio.
 We find that the mass--metallicity relation of our  individually measured sources agrees with that for optical-selected star-forming galaxies at $z\sim0.1$, while metallicities of stacked spectra agree with that of MS galaxies at $z\sim0.78$.
Considering high SFR of individually measured sources, fundamental metallicity relation (FMR) of the IR galaxies is different from that at z$\sim$0.1.
  However, on the mass-metallicity plane, they are consistent with the MS galaxies, highlighting higher SFR of the IR galaxies.
  This suggests the evolutionary path of our infrared galaxies is different from that of MS galaxies.
 A possible physical interpretation includes that the star-formation activities of infrared  galaxies at $z\sim0.88$ in our sample are enhanced by interaction and/or merger of galaxies, but the inflow of metal-poor gas is not yet induced, keeping the metallicity intact.
 \end{abstract}

\section{Introduction}
Stellar mass, star formation rate (SFR), and gas-phase metallicity (hereafter metallicity) of a galaxy are crucial parameters for characterizing a galaxy, and their relations are important for understanding the buildup of galaxies over cosmic time.
\citet{Lequeux79} first found the correlation between the total mass of galaxies and the heavy element abundance (mass--metallicity relation) such that more massive galaxies have higher metallicity.
After that, enormous efforts in studying the mass--metallicity relation in various redshifts have been made at $z\sim0.1$ using SDSS data \citep{Tremonti04, AndrewsMartini13, Zahid14}, at $z<1$ (\cite{Savaglio05, Lara-lopez09, CowieBarger08, Zahid11, Ly14}), at $z\sim1-2$ (\cite{Yabe12, Yabe14, Yabe15, Zahid12, Zahid14, Stott13, Hayashi15,Ly16}), and $z>2$ (\cite{Erb06, Maiolino08, Mannucci09, Hayashi09}). 
The compilation of those studies suggests that the mass--metallicity relation evolves with redshift, in the sense that the metallicity is gradually increasing with decreasing redshift at a given stellar mass \citep{Maiolino08, Zahid13}.

The intrinsic scatter of the mass--metallicity relation and its dependence on SFR has also been investigated mainly in the local Universe \citep{Ellison08a, Yates12, AndrewsMartini13}, while less clear SFR-dependence is found at high redshifts \citep{Yabe12, Yabe14, Yabe15, Steidel14, Sanders15}.
\citet{Mannucci10} has found that the scatter of the mass--metallicity relation for the local galaxies is reduced by accounting for the SFRs. 
They also have suggested that a relationship among the stellar mass, metallicity, and SFR does not evolve for $z<2.5$ galaxies.
The tight surface in this 3-D space is referred as the fundamental metallicity relation (FMR).
It suggests that the evolution of the mass--metallicity relation is due to the apparent shift on this surface with changing SFRs.
\citet{Lara-lopez10} has found a similar relation independently, and many other authors have confirmed the relation in the local Universe \citep{Yates12, AndrewsMartini13,  Lara-lopez13, Perez-Montero09}.

One possibility to explain the formation of the FMR is a balance between pristine gas inflow, metal-rich gas outflow, and star-formation. 
\citet{Dave11} have made $N$-body + smoothed particle hydrodynamic simulations.
They found that a complex interplay of the gas inflow over cosmic time which dilutes metal in interstellar medium and at the same time increases metallicity by supporting star-formation and the outflow which decreases metallicity by transporting the gas from the interstellar medium to the intergalactic region and inhibits the star-formation activity, establishes the FMR with almost no evolution from $z=3$ to $z=0$ (see also \cite{FinlatorDave08, Dave12}).
\citet{Lilly13} have introduced a physically motivated model that predicts the metallicity of the ISM as a function of $M_{\ast}$ and SFR, with infalling and outflowing gas regulating star formation and chemical enrichment in a galaxy. 
The relation given by equation 40 of \citet{Lilly13} closely follows the observed FMR.

However, the existence of the FMR has not been confirmed at various high redshifts. 
Some studies for high redshift galaxies support a non-evolving FMR \citep{Belli13, Stott13, Cullen14, Yabe15}, whereas other works insist the FMR evolves at $z>1.5$ \citep{Wuyts14, Zahid14, Sanders15}.

Studies of the extragalactic background have suggested that at least one third (or half) of the luminous energy generated by stars is hidden by dust and reprocessed into the infrared emission at a redshift of $z\sim1$ \citep{Puget96, Lagache99, Stecker06, Franceschini08}, indicating that significant star-formation is occurring in dusty regions.
Thus, to understand the true star formation history (SFH), it is crucial to investigate infrared (IR) bright galaxies.
Most of the previous studies of the relationship between stellar mass, metallicity, and SFR are for optical (or near-IR) selected galaxies. 
Recently, the mass--metallicity relation of IR galaxies has been studied in the local Universe. 
Luminous and ultra luminous infrared galaxies (LIRGs and ULIRGs) in the local Universe are found to have lower metallicity at a similar stellar mass compared with optically selected star-forming galaxies \citep{Rupke08, Rodrigues08, Kilerci-Eser14}.
On the other hand, a study of metallicity of $Herschel$ selected far-IR galaxies at $z>1$ by \citet{Roseboom12} do not show any clear metallicity deficit compared with local optical selected galaxies (see also \cite{Silverman15}).
That is, there is no general consensus yet.
It is an urgent task to investigate the mass-metallicity relation at high-$z$, using infrared light, which carries the bulk of the star-formation activity.
In this work, we investigate AKARI mid-IR selected galaxies at $z\sim1$ in the north ecliptic pole (NEP) region to examine the mass--metallicity relation and its SFR dependence to explore cosmic SFH. 
Throughout this paper, cosmological parameters of $H_0$ = 70 km~s$^{-1}$~Mpc$^{-1}$, $\Omega_{\Lambda}$ = 0.7, and $\Omega_{m}$ = 0.3 are adopted.
We assume a \citet{Chabrier03} initial mass function (IMF).

\section{Sample Selection, Observation and Reduction}

\subsection{Photometric Data in the AKARI NEP-Deep field}
\label{Photometric Data in the AKARI NEP-Deep field}
Our targets were selected from AKARI NEP-Deep field catalog by \citet{Murata13}.
The AKARI NEP-Deep field survey carried out deep photometry over $\sim$0.5 sq.deg in 9 continuous bands of 2, 3, 4, 7, 9, 11, 15, 18, and 24 $\mu$m with a 5 $\sigma$ photometric sensitivity of 13, 10, 12, 34, 38, 64, 98, 105, and 266 $\mu$Jy, respectively.
Many astronomical satellites have accumulated deep exposures that cover the NEP-Deep region, X-ray observations by $Chandra$ \citep{Krumpe15}, UV observations by GALEX (Program GI4-057001-AKARI-NEP, P.I. M. Malkan), and far-IR observations by $Herschel$ (Program OT1-sserj01-1, P.I. S. Serjeant; Pearson et al. in prep.).
Optical and near-IR photometric data were also taken with ground-based telescopes \citep{Wada08, Oi14}.
We first selected AKARI mid-IR sources detected in more than one bands among the 11, 15, and 18 $\mu$m filters. 
Since a star-forming galaxy shows a strong Polycyclic Aromatic Hydrocarbons (PAH) emission feature at 7.7 $\mu$m, which enters into the mid-IR bands at $z\sim1$, the mid-infrared selection is sensitive to star-forming IR galaxies at that redshift. 
We used photometric redshifts ($z_p$) derived from the Spectral Energy Distribution (SED) fitting with photometric data covering from optical to near-IR \citep{Oi14}. 
For the $z_p$ calculation, 62 galaxy templates \citep{Coleman80, Kinney96} and 154 star templates \citep{Pickles98} are fitted with extinction laws \citep{Calzetti00, Prevot84} with 2175 $\AA$ UV bump and 8 reddening values ($E(B-V)$) from 0.0 to 0.5 using $Le~Phare$ code \citep{Ilbert06, Ilbert09, Arnouts07}.
The accuracy of this $z_p$ is $\sigma_{\Delta z}/(1+z) = 0.032$ and the catastrophic error ($\Delta z/(1+z)>0.15$) rate is $\eta=5.8$\% at $z<1$.
Since our aim is to observe H$\alpha$ and [N \emissiontype{II}] emission lines to measure metallicities in $J$-long band (1.11 -- 1.35 $\mu$m) of Subaru/FMOS with high resolution mode, we extracted sources from the AKARI mid-IR sample with reliable $z_p = 0.637 - 1.123$ determined with five or more photometric data. 

In order to increase the size of the target list, we added in $\sim$ 500 AKARI mid-IR sources with spectroscopic redshifts ($z_s$) consistent to be within 0.691 -- 1.057 within 1 $\sigma$ taken with MMT/Hectospec, WIYN/HYDRA \citep{Shim13}, and Keck/DEIMOS (Takagi in prep).
As a result, 1528 sources are included in our candidate target list.

Next, we estimated expected H$\alpha$ emission line fluxes from IR luminosity ($L_{\rm IR}=\int^{1000\mu {\rm m}}_{8\mu {\rm m}}L_{\lambda}dL_{\lambda}$) by using the $Le~Phare$ code.
We fit model templates to the observed SEDs from optical $u^{*}$-band to mid-IR 24 $\mu$m by standard $\chi^{2}$ minimization.
We used stellar templates of \citet{BruzualCharlot03} in optical and near-IR wavelength and dust emission templates of \citet{DaleHelou02}, \citet{CharyElbaz01}, and \citet{Lagache03} in mid-IR (and far-IR) wavelength.
Exponentially decreasing star formation models (SFR $\propto e^{-t/\tau}$) with $\tau$ = 0.1, 0.3, 1, 2, 3, 5, 10, 15, 30 Gyrs were adopted.
We applied the extinction law of \citet{Calzetti00} with 2175$\AA$ bump.
Since \citet{Roseboom12} found that a range of $E(B-V)$ for infrared galaxies detected by $Herschel$ at $z\sim 1$ is 0.0 -- 1.0, we allowed the $E(B-V)$ to vary from 0.0 to 1.0.
Stellar population age ranged from 3 Myrs to 13 Gyrs.
The $z_p$ from \citet{Oi14} was fixed during determining the physical parameters of the galaxies.
The intrinsic H$\alpha$ luminosity ($L_{\rm H\alpha, int}$) was calculated from the $L_{\rm IR}$ using the conversion in \citet{Kennicutt98}.
Then, considering the dust attenuation and the extinction difference between ionized gas and stellar components of 0.44 times \citep{Calzetti97}, we predicted H$\alpha$ emission line flux.
We prioritized targets in depending order of their expected H$\alpha$ emission flux.

\subsection{Observation and Data reduction} 
The observations were carried out with the Fiber Multi Object Spectrograph (FMOS; \cite{Kimura10}) on the Subaru Telescope on June 20 and 21, 2012.
The high-resolution (HR) mode (R $\sim$ 2200) $J$-long band (1.11 -- 1.35 $\mu$m) gives better 
line sensitivity in the wavelength range ($0.4\times10^{-16}$~erg~s$^{-1}$~cm$^{-2}$) 
than in the other bands of HR mode and low-resolution (LR) mode ($1.0\times10^{-16}$~erg~s$^{-1}$~cm$^{-2}$).
The typical seeing monitored in the $R$-band was \timeform{0.9"} -- \timeform{1.0"}.

The fiber configuration design for the targets in each FMOS field of view was prepared by using the FMOS fiber allocation software which
semi-automatically optimizes the target priorities, with some restrictions such as the fiber motion or physical size of fiber tip.
We mainly used Cross Beam Switching (CBS) mode, in which each fiber observed a source (position A) and sky (position B) alternately, and two fibers were allocated for the same target one after the other.
Then, we allocated the rest of the fibers manually to sources with Normal Beam Switching (NBS) mode, in which only one fiber was allocated to one source and observed a source (position A) and sky (position B) alternatively.
In order to share the FMOS fibers effectively we observed sources for other science motivations with lower priority.
Consequently, 354 sources out of 1528 sources in our target list were allocated to fibers.
We also selected several F, G, or K-type stars using  $g^{'}-r^{'}$, $J-H$, or $H-K_{s}$ colors 
to provide flux calibrators, observed at the same time as the scientific targets.
We observed five fields of view in the AKARI NEP-Deep field centered at ($\alpha$, $\delta$) =  
(\timeform{17h55m53.5s}, \timeform{+66D45'25.7"}), 
(\timeform{17h54m0.9s}, \timeform{+66D29'32.9"}), 
(\timeform{17h56m24.90s}, \timeform{+66D21'19.8"}), 
(\timeform{17h55m0.8s}, \timeform{+66D24'15.9"}), and
(\timeform{17h54m33.8s}, \timeform{+66D38'33.0''}).
Many objects are located in overlapped area of more than one field of view, and some of them were observed multiple times in different field of view.
One exposure time was 15 minutes at position A.
Then we moved to position B and observed for another 15 minutes.
The total exposure times ranged from 15 -- 255 minutes, with an average of 60 minutes.
 
The obtained data were reduced with the FMOS pipeline {\tt FIBRE-PAC} \citep{Iwamuro12}, which
The {\tt FIBRE-PAC} can reduce data taken in a fiber configuration in a field with CBS or NBS reduction mode.
Since we mixed CBS mode and NBS mode observations in one fiber configuration, we reduced all data twice with each mode. 
The basic processes of {\tt FIBRE-PAC} are as follows;  
(i) sky subtraction by the subtraction of a set of exposures $A - B$,
(ii) distortion correction of the sky subtracted 2-D spectra,
(iii) combining spectra of each target taken in different exposures with the same configuration,
 (iv) wavelength calibration using Th-Ar lamp frames,
(v) flux calibration using the spectra of standard stars.
For the CBS mode, there is an additional process to combine spectra of each target obtained by pair fibers.
The uncertainty associated with the wavelength calibration was $<1.2$~\AA~\citep{Iwamuro12}.
Readers are referred to \citet{Tonegawa15} for a more automatic line detection algorithm from FMOS data.

\begin{figure}[htbp]
\begin{center}
  \includegraphics[width=100mm]{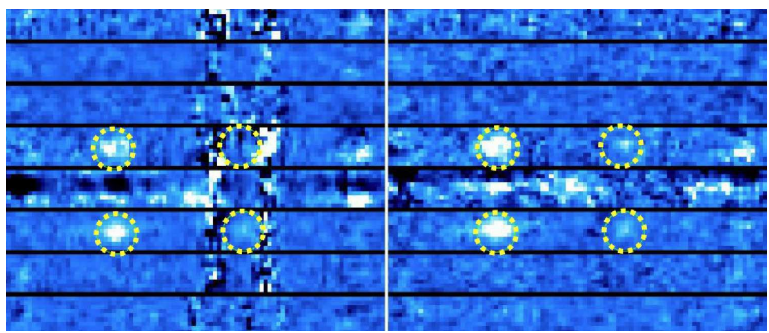}
\caption{Examples of 2-D spectra before and after the flux calibration. 
Wavelength increases along the $x$-direction. 
A spectrum is spread in the $y$-direction across 9 rows, each of which is separated by black horizontals. A spectrum after flux calibration (left panel) shows the destroyed emission lines (yellow circles) caused by the division process by a spectrum of a standard star with low $S/N$ region due to the OH air glow masks. On the other hand, the emission lines are clearly seen in the spectrum before the flux calibration (right panel).}
\label{fig:LineDestroyImage}
\end{center}
  \end{figure}

For obtaining the relative flux calibrated spectra, we slightly modified the {\tt FIBRE-PAC} pipeline.
In the near-IR wavelength range, there are many OH night air glow emission lines.
FMOS has OH air glow masks printed on the surface of the mask mirror directly (physically) 
at the corresponding locations of the stronger air glow emission lines.
The intensity of spectra at the masked pixels decreases, and the signal-to-noise ratio ($S/N$) worsens.
Since pixel -- wavelength relations are slightly different for each fiber, the masked wavelengths are slightly different between objects. 
{\tt FIBRE-PAC} aligns pixels (not wavelength) of spectra for a science target and a standard star and carries out the flux calibration by division of the spectra.
Although this helps to suppress noise, flux calibration is made at a slightly different wavelengths.
In order to calibrate flux with the same wavelength, we modified the {\tt FIBRE-PAC} and aligned spectra of a science target and a standard star with wavelength instead of pixel.
Due to the offset of masked wavelengths between an object and a standard star, a flux-calibrated spectrum has wider noisy regions than normal.
The wider noisy regions increase the chance to lose important emission lines, and inhibit the measurement of the line fluxes or even identifying a line feature (left panel of Figure \ref{fig:LineDestroyImage}).
To overcome this difficulty, we again modified the {\tt FIBRE-PAC}, to output a spectrum before the flux calibration.
Although these spectra are also effected by the OH air glow masks, they are not as noisy as spectra after the flux calibration (right panel of Figure \ref{fig:LineDestroyImage}).
Thus, we used these less noisy spectra for emission identifications.

As mentioned above, since some sources were observed multiple times with different observation modes and/or different field of view, we combined spectra for each object, which were reduced separately, and finally created average spectra weighted by exposure time.
From the weighted average spectra before flux calibration, we found emission features from 75 sources out of 354 observed targets.
In order to identify whether these features are H$\alpha$ emission lines, we used information of other emission lines.
[N \emissiontype{II}]$\lambda$6584 emission line were detected in 21 FMOS spectra. 
For 7 sources out of the remaining 54 sources, other emission line such as [O \emissiontype{II}]$\lambda$3727, H$\beta$, and [O \emissiontype{III}]$\lambda$5007 were detected in already existing optical spectra (e.g., \cite{Shim13}, Takagi in prep.).
Therefore, we considered the 28 sources as secure H$\alpha$-detected sources. 
Emission lines of one object (AKARI ID = 61012430) were identified as [O \emissiontype{III}]$\lambda \lambda$4959,5007, and emission lines of two objects (AKARI ID = 61020689 and 61022567) were [O \emissiontype{II}]$\lambda \lambda$3727,3729. 
The emission lines of the rest of 44 sources could not be identified. 
The detection rate of the secure H$\alpha$ is 7.9 \% (= 28/354), and when we assume that the single emission lines from the 44 sources are all H$\alpha$, then the detection rates of H$\alpha$ is 20.3 \% (= 72/354).

\begin{figure}[htbp]
\begin{center}
  \includegraphics[width=110mm]{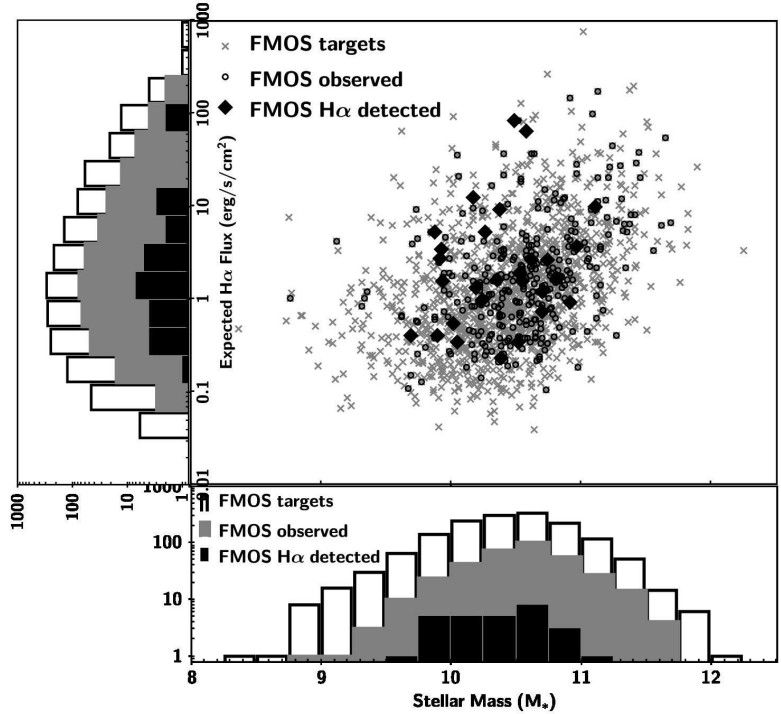}
\caption{Distribution of stellar mass and expected H$\alpha$ flux for the target sample (1528 sources; crosses), FMOS observed sample (354 sources; open circles), and secure H$\alpha$ detected sample (28 sources; filled diamonds). In the bottom and left-hand panels, the histograms of the stellar mass and expected H$\alpha$ flux are shown for each sample in open, gray-filled, and black-filled, respectively.}
\label{fig:MstarfHatargets}
\end{center}
  \end{figure}

In Figure \ref{fig:MstarfHatargets}, the stellar masses of our 1528 targets, our FMOS-observed galaxies and of our H$\alpha$-detected sub-sample 
are plotted against their expected H$\alpha$ emission fluxes.
Although we set a priority on targets with strong expected H$\alpha$ emission, 
the fiber-allocated sources and H$\alpha$-detected sources are not strongly biased toward higher H$\alpha$ flux, or larger stellar mass.
This is probably because of the restrictions of fiber allocation.
Basic information for the secure H$\alpha$ detected sources is summarized in Table \ref{tb:basic}.
The information of all the single H$\alpha$-detected sources is summarized in the Appendix.

\begin{table}[t]
\begin{center}
\caption{{Information of secure H$\alpha$ detected sources. \hfil\hfill}}
\label{tb:basic}
\small
\begin{tabular}{lccccccc}
\hline
  \multicolumn{1}{c}{Object} &
  \multicolumn{1}{c}{RA [deg]} &
  \multicolumn{1}{c}{DEC [deg]} &
  \multicolumn{1}{c}{$f_{100\mu {\rm m}}$[mJy]} &
  \multicolumn{1}{c}{$f_{160\mu {\rm m}}$[mJy]} &
  \multicolumn{1}{c}{$f_{250\mu {\rm m}}$[mJy]} &
  \multicolumn{1}{c}{$f_{350\mu {\rm m}}$[mJy]} &
  \multicolumn{1}{c}{$f_{500\mu {\rm m}}$[mJy]} \\
  \multicolumn{1}{c}{(1)} &
  \multicolumn{1}{c}{(2)} &
  \multicolumn{1}{c}{(3)} &
  \multicolumn{1}{c}{(4)} &
  \multicolumn{1}{c}{(5)} &
  \multicolumn{1}{c}{(6)} &
  \multicolumn{1}{c}{(7)} &
  \multicolumn{1}{c}{(8)}\\
  \hline
   61007260         & 269.29566 & 66.33775 & 20.51 $\pm$ 7.25 & 32.89 $\pm$ 11.47 &   26.54 $\pm$  6.30 & 32.74 $\pm$24.74 &               ...                \\
   61010028         & 268.78815 & 66.42292 & 11.19 $\pm$ 2.48 &                ...             &                ...              &               ...             & 127.69 $\pm$ 88.02 \\
   61010435         & 268.52110 & 66.43493 &   4.94 $\pm$ 4.45 & 21.09 $\pm$ 21.07 &   19.40 $\pm$14.13 & 34.72 $\pm$24.59 &               ...                \\
   61010515         & 268.76926 & 66.43767 & 12.86 $\pm$ 2.27 &                ...             &   20.83 $\pm$  8.26 &               ...             &               ...                \\
   61011420         & 268.90074 & 66.46313 & 12.86 $\pm$ 0.71 &                ...             &   28.91 $\pm$  7.30 & 21.96 $\pm$  5.90 &   16.21 $\pm$ 10.94 \\
   61011677         & 268.89527 & 66.47111 &            ...               & 32.07 $\pm$    8.77 &  22.53 $\pm$12.19 & 19.01 $\pm$  8.10 &   13.00 $\pm$ 12.01 \\
   61012118         & 269.10461 & 66.48428 &   8.96 $\pm$ 3.03 &                ...             &   10.41 $\pm$  4.79 & 14.61 $\pm$  5.16 &   18.58 $\pm$   9.21 \\
   61012132         & 268.12720 & 66.48165 & 21.30 $\pm$ 7.11 & 13.33 $\pm$   7.24 &               ...               &               ...             &               ...                \\
   61012133         & 269.04122 & 66.48479 &            ...               &                ...             &   13.15 $\pm$  4.97 & 18.11 $\pm$10.12 &               ...                \\
   61012385         & 268.97977 & 66.49096 &   5.84 $\pm$ 5.70 &                ...             &               ...               &               ...             &               ...                \\
   61013116         & 268.44526 & 66.51066 &   7.48 $\pm$ 2.45 & 55.41 $\pm$ 11.58 &   35.62 $\pm$  8.70 & 36.60 $\pm$  8.54 &   13.61 $\pm$   6.48  \\
   61013936         & 268.40091 & 66.53157 &            ...               &                ...             &   13.02 $\pm$  4.28 & 10.87 $\pm$  5.98 &   19.47 $\pm$   6.72 \\
   61014553         & 268.32456 & 66.54723 &            ...               &                ...             &               ...               &               ...             &               ...                \\
   61016374         & 269.17252 & 66.59399 &            ...               &                ...             & 111.75 $\pm$77.07 &               ...             &               ...                \\
   61017060         & 268.44108 & 66.61139 &            ...               &                ...             &               ...               &               ...             &               ...                \\
   61017881         & 268.31424 & 66.63291 &            ...               & 43.46 $\pm$ 12.31 &   40.38 $\pm$15.32 & 14.40 $\pm$  5.96 &   35.74 $\pm$ 27.14 \\
   61018324$^a$ & 268.81247 & 66.64760 &            ...               &                ...             &               ...               &   9.75 $\pm$  4.17 &   12.76 $\pm$ 10.51 \\
   61019568         & 268.49825 & 66.68160 &   8.47 $\pm$ 4.98 & 18.39 $\pm$ 12.56 &               ...               &               ...             &               ...                \\
   61020367$^a$ & 269.18132 & 66.70350 &            ...               &                ...             &   11.80 $\pm$  7.17 &               ...              &               ...                \\
   61020444         & 268.47018 & 66.70364 & 10.77 $\pm$ 1.94 & 21.58 $\pm$ 13.83 &   28.80 $\pm$  5.00 & 33.04 $\pm$  9.90 &               ...                \\
   61021272         & 268.35996 & 66.72962 & 18.64 $\pm$ 6.90 & 15.24 $\pm$   5.76 &   13.56 $\pm$  4.00 & 20.21 $\pm$  3.98 &   19.63 $\pm$   5.57 \\
   61022683         & 268.39410 & 66.77837 &            ...               &   9.57 $\pm$   4.87 &   14.50 $\pm$  7.70 & 23.38 $\pm$15.01 &               ...                \\
   61022934         & 268.36473 & 66.78714 &   5.95 $\pm$ 4.46 &                ...             &   16.22 $\pm$  9.99 & 12.10 $\pm$10.90 &   12.74 $\pm$   7.77 \\
   61023221         & 268.53500 & 66.80033 &            ...               &                ...             &               ...               & 23.33 $\pm$14.37 &   37.11 $\pm$ 24.27 \\
   61023846         & 269.18879 & 66.82412 &            ...               &                ...             &   14.60 $\pm$  8.77 &              ...              &   17.09 $\pm$   9.09 \\
   61024055         & 268.40068 & 66.83113 &            ...               & 20.69 $\pm$   8.86 &   33.46 $\pm$14.20 & 17.76 $\pm$  8.51 &   18.15 $\pm$   9.93 \\
   61024136         & 268.73772 & 66.83440 & 10.05 $\pm$ 3.81 & 32.27 $\pm$ 11.54 &   33.99 $\pm$  7.32 & 24.60 $\pm$  6.81 &   30.62 $\pm$   6.00 \\
   61024723         & 268.73494 & 66.86060 &            ...               & 21.51 $\pm$ 17.33 &   44.69 $\pm$  5.94 & 48.70 $\pm$  8.42 &   29.80 $\pm$ 11.60 \\
\hline
\end{tabular}
\end{center}
(1): AKARI ID from \citet{Murata13}. (2) and (3): position of source. (4)--(8): $Herschel$ photometric data.\\
$^a$ Sources detected by $Chandra$.\\
\end{table}

\subsection {Spectral line fitting}
\label{sec:LineFitting}

\begin{figure}[htbp]
  \begin{center}
\includegraphics[width=50mm,angle=-90]{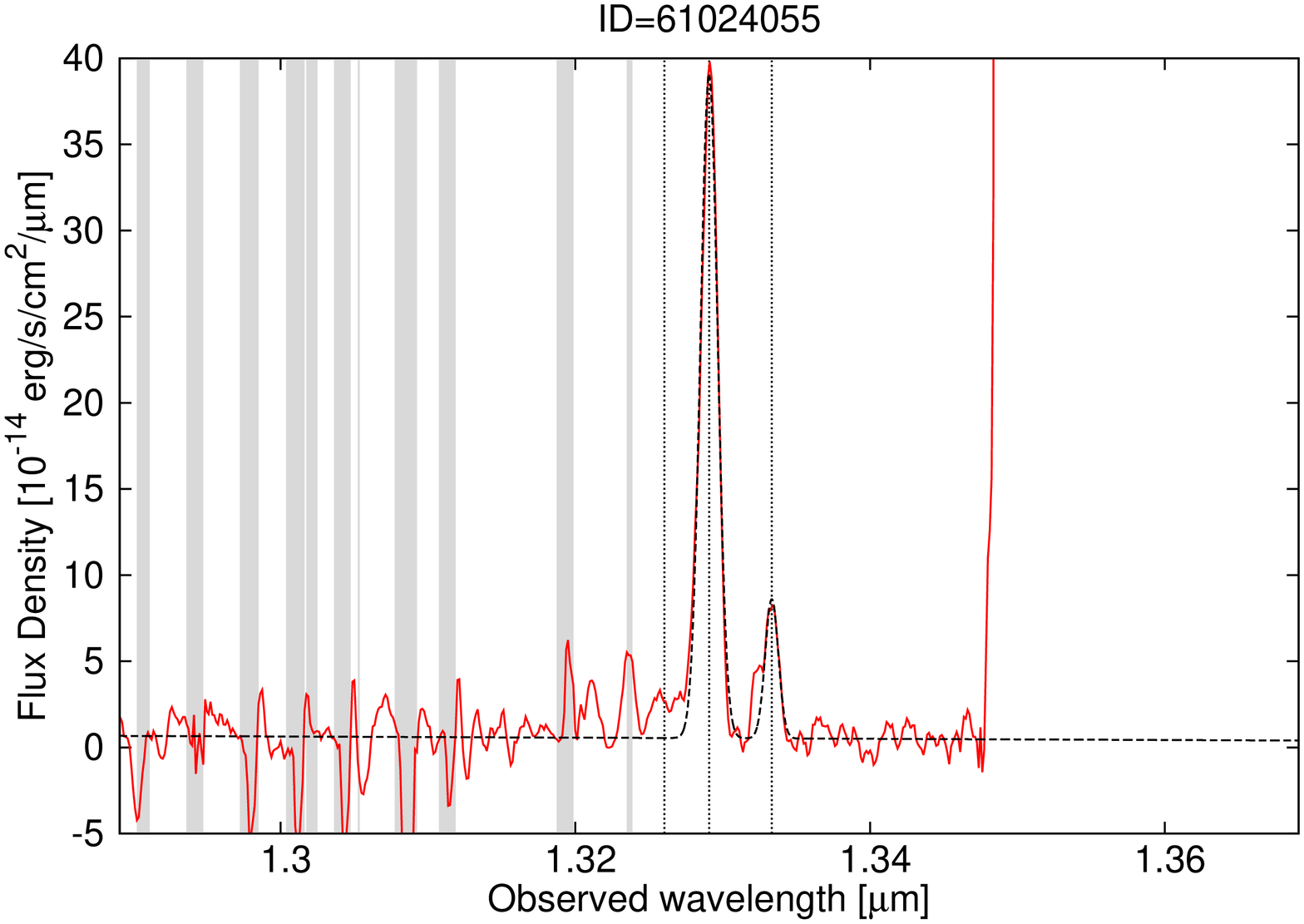}
\includegraphics[width=50mm,angle=-90]{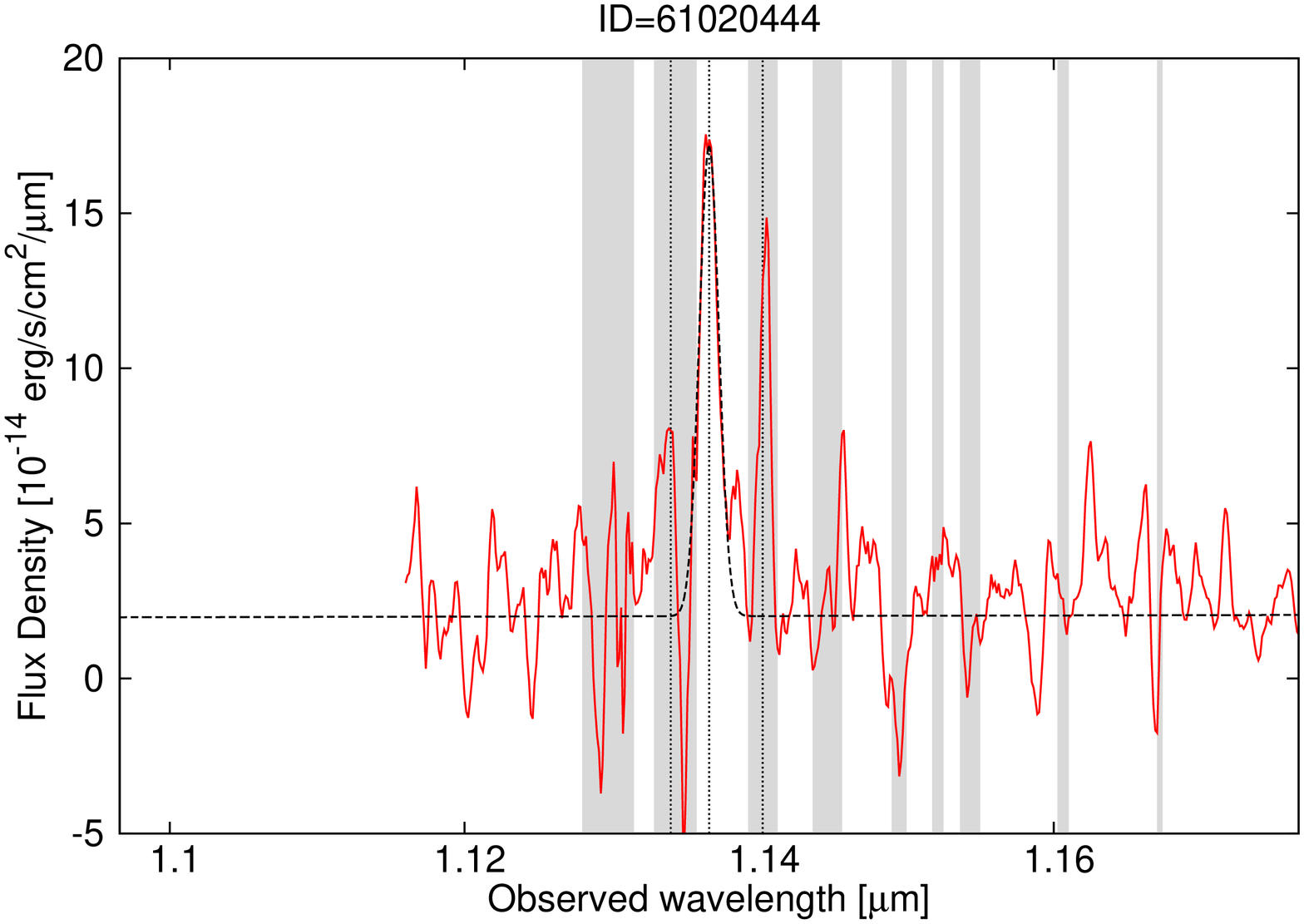}
\includegraphics[width=50mm,angle=-90]{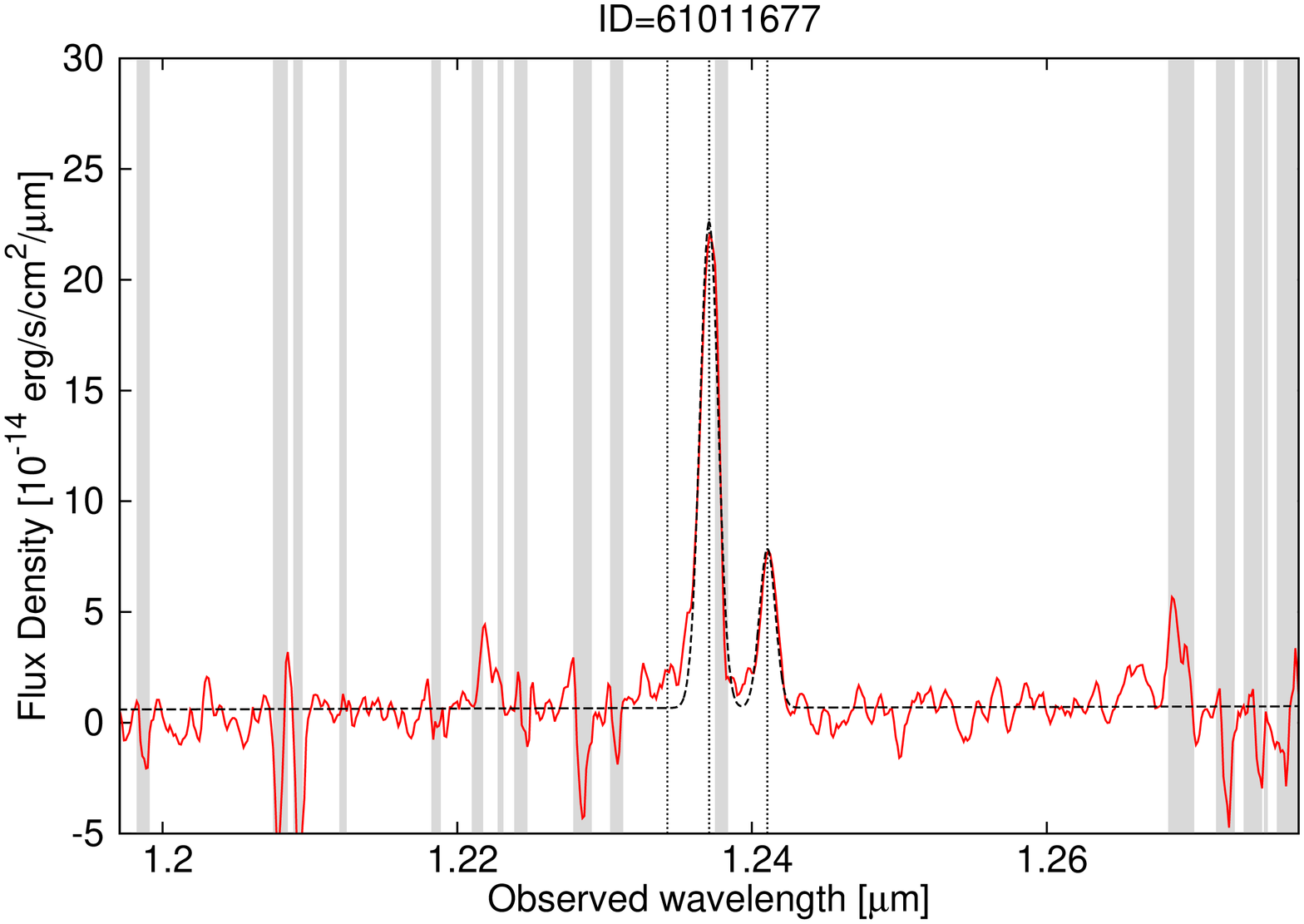}
\includegraphics[width=50mm,angle=-90]{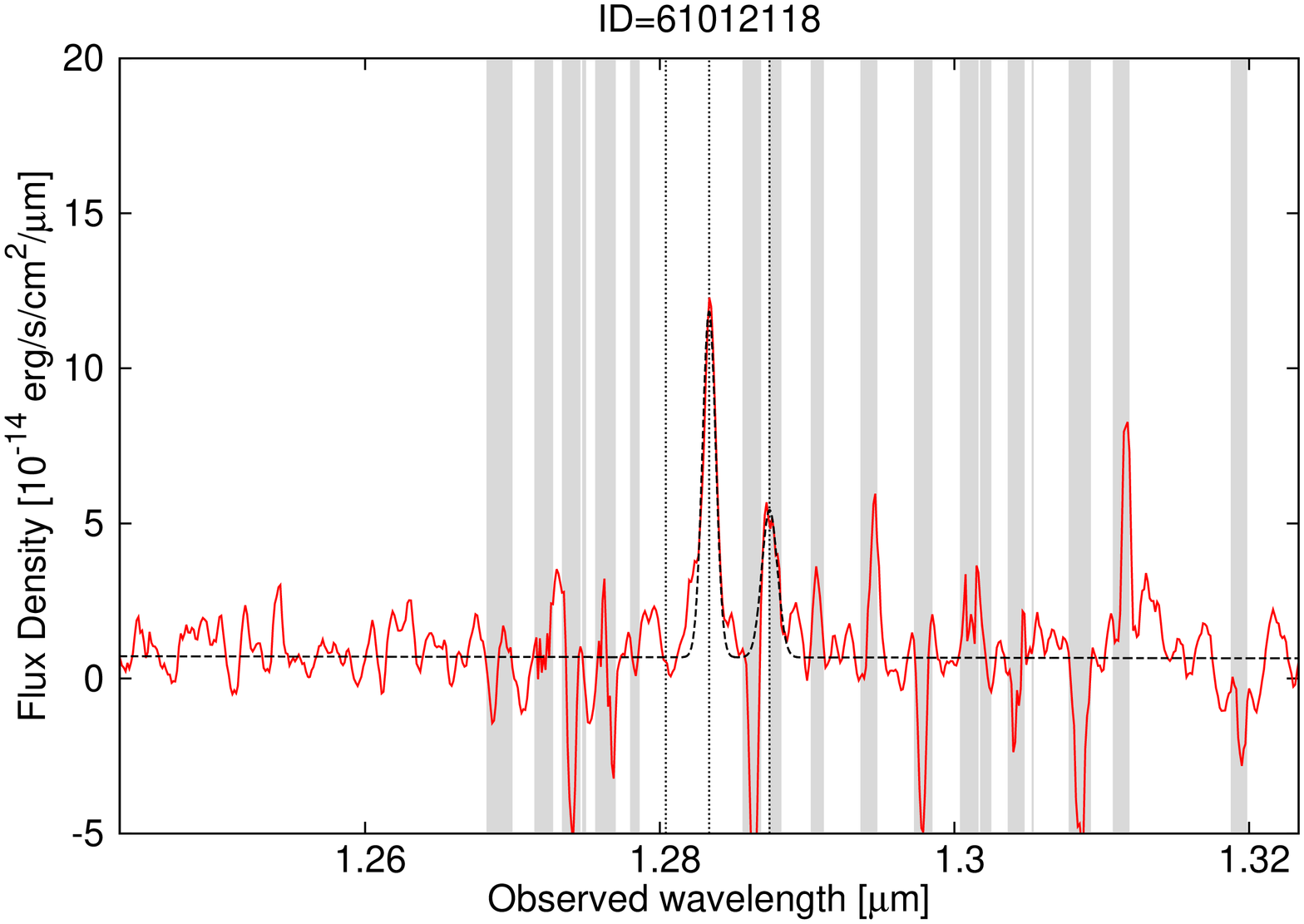}
    \caption{Examples of FMOS spectra. Emission line(s) and continuum fitting results are overlaid with dashed lines. Gray shadows represent OH mask regions. Three vertical lines in each panel indicates the position of [N \emissiontype{II}]$\lambda$6548, H$\alpha$, and [N \emissiontype{II}]$\lambda$6584 from left to right. (Top left): an example of good H$\alpha$ and [N \emissiontype{II}]$\lambda$6584 emission lines. (Top right): good single line detection. With existing spectroscopic redshift, the detection line is confirmed as H$\alpha$. (Bottom left): H$\alpha$ and [N \emissiontype{II}] emission lines are detected but half of H$\alpha$ is destroyed. The other half of the H$\alpha$ emission line is used for the fitting. (Bottom right): [N \emissiontype{II}] line is almost completely destroyed although somehow the peak is detected. The center and width of the [N \emissiontype{II}] are assumed from the H$\alpha$ fitting result.}
    \label{fig:MZRupke}
  \end{center}
\end{figure}

\begin{figure}[htbp]
\begin{center}
  \includegraphics[width=60mm]{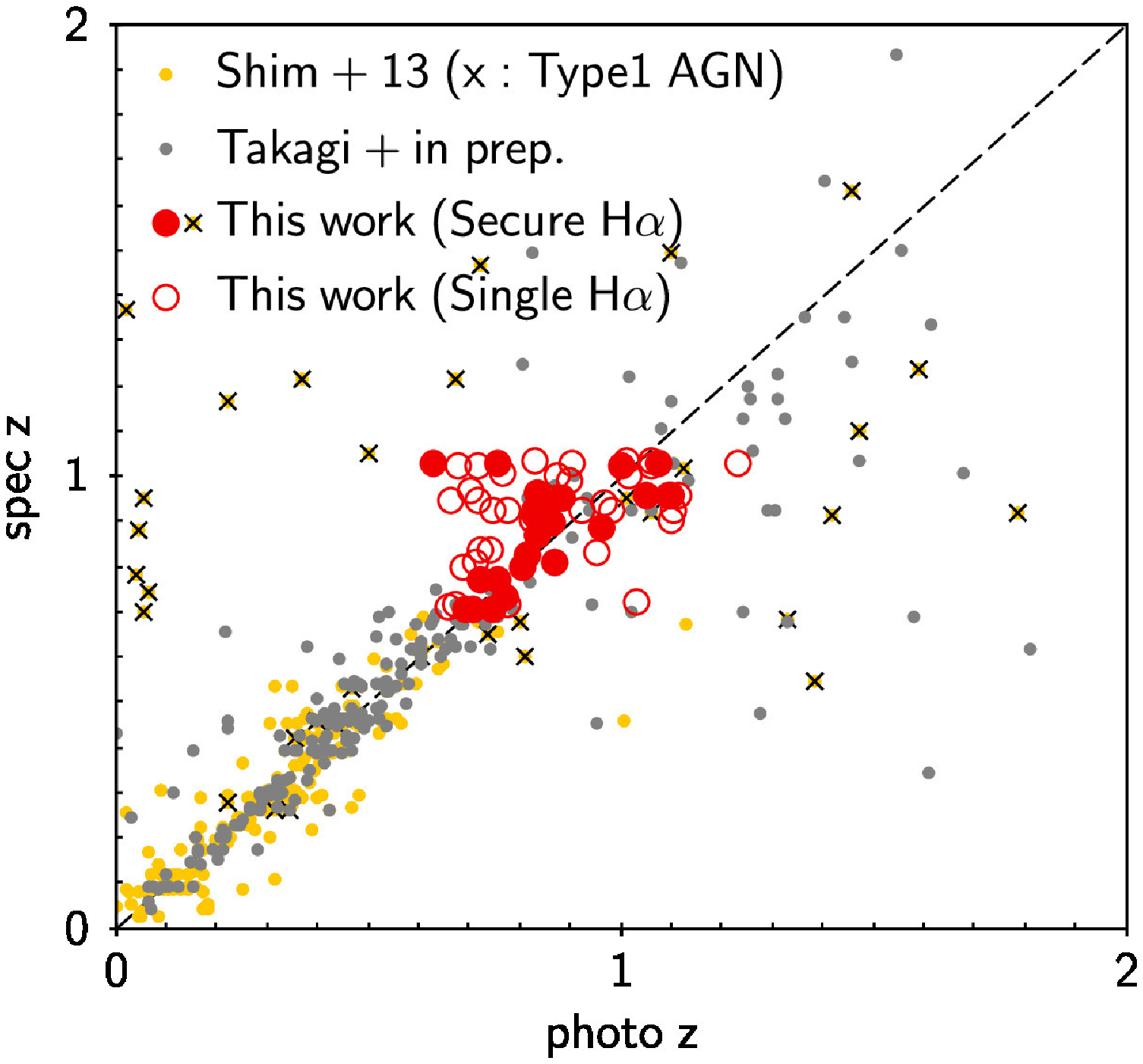} 
  \includegraphics[width=65mm]{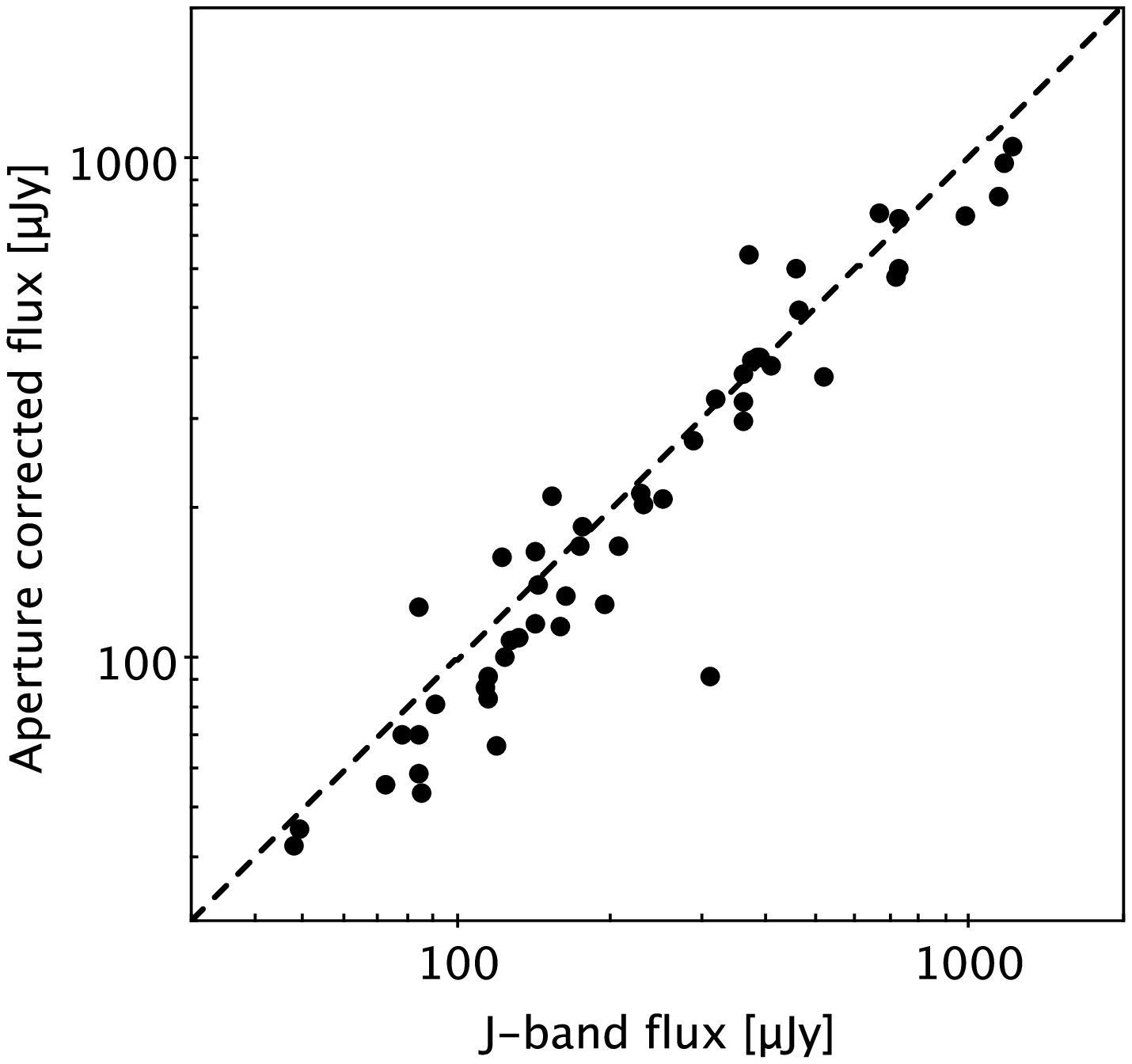}
  \includegraphics[width=75mm]{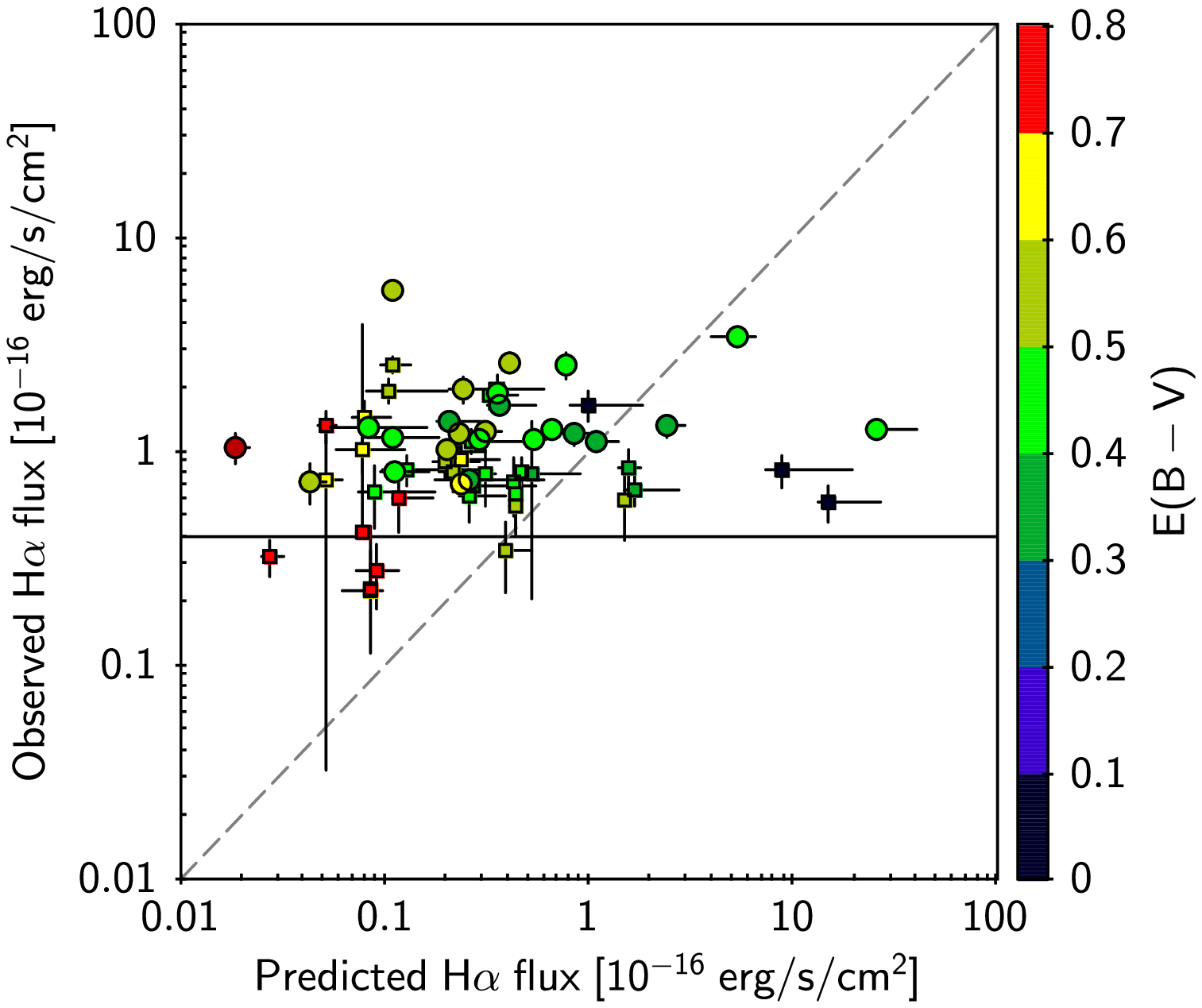}
\caption{(Top left) Comparison between photometric redshift and spectroscopic redshift at $0<z<2$. Red filled and opened circles represent the secure H$\alpha$ detected sources and single emission line detected sources with an assumption that the emission line is H$\alpha$ of this work, respectively. 
The existing spectroscopic data (\cite{Shim13}; yellow, Takagi+in prep.; gray) are also plotted.
The crosses show sources with type1 AGN signs. 
(Top right) The aperture corrected flux compared with CFHT/WIRCam $J$-band flux. 54 sources whose continuum emission are clearly detected are plotted. 
(Bottom) Comparison of the predicted H$\alpha$ flux from $L_{\rm IR}$ and $E(B-V)$ by SED fitting ($\S$ \ref{Photometric Data in the AKARI NEP-Deep field}) and the observed H$\alpha$ flux in $10^{-16}~{\rm erg}~{\rm s}^{-1}~{\rm cm}^{-2}$ for line detected sources. The circles and squares are for secure H$\alpha$ detected sources and for the single H$\alpha$ detected sources, respectively. 
The color-code is their $E(B-V)$.
The average detection limit of our observation is at 4.0 $\times 10^{-17}$ erg s$^{-1}~{\rm cm}^{-2}$ (black horizontal line). 
}  
  \label{fig:ObsVSPred}
\end{center}
  \end{figure}

In order to measure a spectroscopic redshift and fluxes of emission lines, we fit models to 1-D spectra extracted from the absolute flux-calibrated 2-D spectra.
We fitted Gaussians to the emission lines, and a linear continuum.
For fitting the continuum, we used data at three main wavelength ranges of 1.170 -- 1.194$\mu$m, 1.243 -- 1.266$\mu$m, and 1.325 -- 1.340$\mu$m, where there is no strong OH emission.
If there were emission lines in those ranges, they were excluded from the fitting.
For the emission line fitting, when both H$\alpha$ and [N \emissiontype{II}]$\lambda$6584 lines were detected, we fit the both lines simultaneously with two Gaussian functions with a fixed peak wavelength ratio of (0.6563$\mu$m / 0.6584$\mu$m) $\times$ (1+$z$), where $z$ is a free parameter.
We show four example spectra in Figure \ref{fig:MZRupke}.
When an emission line shifted into a OH masked region and was partially destroyed by the mask, we did not use the destroyed data for the fitting.
When both H$\alpha$ and [N \emissiontype{II}]$\lambda$6584 lines were detected, but one line was severely affected, we assumed that the line width was the same for the both lines and used a line width derived from the other line.
Two sources (AKARI ID = 61010028 and 61012132) showed their [N \emissiontype{II}]$\lambda$6584 lines in their spectra for the line identification but could not be fit due to the OH airglow. 
For sources whose [N \emissiontype{II}] is detected with $<$3$\sigma$, we measured 3$\sigma$ upper limit flux, where a 1$\sigma$ is defined by a standard deviation of their continuum.
For the 44 single emission line detected sources, we could fit the emission line of 36 sources.
We measured their $z_s$ with the assumption that the line is H$\alpha$ emission.
Obtained spectroscopic redshifts from the fitting are compared with their $z_p$ (left panel of Figure \ref{fig:ObsVSPred}).
The spectroscopic redshifts by FMOS agree with their $z_p$ with $\langle \Delta z\rangle=0.02$ and $\sigma=0.09$. 
The obtained redshifts are summarized in Table \ref{tb:detail} and in the Appendix.

The measured flux is affected by fiber loss due to the seeing conditions and the extended nature of our galaxies. 
Although the aperture effect on the metallicity using an emission line ratio is small, the effect is critical to SFR and dust attenuation.
To correct for the aperture effect, we estimated a correction factor by using our CFHT/WIRCam $J$-band image \citep{Oi14} for each galaxy.
We convolved the $J$-band image to the typical seeing size of FMOS observation (\timeform{0.95"}), and measured 
the flux from the whole galaxy (FLUX\_AUTO), and also 
that within \timeform{1.2"} diameter aperture (FLUX\_APER1.2) using SExtractor \citep{BertinArnouts96}.
The conversion factor for each galaxy is defined as FLUX\_AUTO / FLUX\_APER1.2. 
The aperture corrected H$\alpha$ and [N \emissiontype{II}] line fluxes (or [N \emissiontype{II}] 3$\sigma$ upper limit fluxes) using the correction factor are reported in Table \ref{tb:detail}.
We also estimated the conversion factors using the existing $u^{\ast}$ and $g^{'}$ images which are consistent with the rest frame UV ones for sources at around $z = 1$ because the rest-frame UV wavelength could be thought as a better tracer of the size of integrated star-forming regions that emits the H$\alpha$ emission.
The conversion factors in the $u^{\ast}$ and $g^{'}$ are both systematically larger than that  of $J$-band by 1.1 -- 1.3.
The differences can be naturally explained by strong dust attenuation effect in the central region at the shorter wavelength.
Thus, in this work, we use the conversion factor measured with the observed $J$-band image with assumption that H$\alpha$ emitting area is spread over the entire galaxy.

To estimate the uncertainty of the aperture-corrected flux, we used 54 continuum-dominated sources in our observed sample including standard stars.
For these sources, we measured continuum flux at 1.253~$\mu$m from our FMOS spectra corresponding to the central wavelength of the WIRCam $J$-band filter, and corrected for the aperture effect.
In the right panel of Figure \ref{fig:ObsVSPred}, we compare the aperture-corrected FMOS flux with their $J$-band flux.
The aperture-corrected fluxes agree well with the $J$-band fluxes, with a scatter is $\sim$ 25\%.
This error includes uncertainty in the FMOS fiber location ($\lesssim 0.2$~arcsec; \cite{Kimura10}).

In the bottom panel of Figure \ref{fig:ObsVSPred}, we compare the predicted and measured H$\alpha$ fluxes. Although a detailed comparison is difficult due to the small sample size and the detection limit of our observation ($4.0\times10^{-17}$~erg~s$^{-1}$~cm$^{-2}$), the figure suggests dusty galaxies with a larger $E(B-V)$ have H$\alpha$ flux underestimated. This could reflect difficulty in estimating extinction in line emitting regions using that of the continuum emitting regions. For galaxies with $E(B-V)<0.5$, the predicted and observed flux agree better.

\subsection{AGN removal}
\label{AGNremoval}
Although we will measure metallicity using the [N \emissiontype{II}]/H$\alpha$ line ratio in next section, 
this ratio is also sensitive to the presence of an active galactic nucleus (AGN).
Thus, we need to exclude AGN-affected galaxies from our sample before we use the line ratio as an indicator of metallicity. 
First, we checked for X-ray detections.
The $Chandra$ X-ray satellite carried out observation of $\sim$0.34~deg$^2$ in AKARI NEP-Deep field \citep{Krumpe15}.
We cross-matched our sample with the X-ray catalog within \timeform{1''} radius searching area corresponding to the pointing accuracy size of the X-ray catalog.
We found two $Chandra$ sources which are likely to be AGNs.
Next, we checked emission line widths of H$\alpha$ and [N \emissiontype{II}]$\lambda$6584 in our FMOS spectra, and found three sources showing significantly broad H$\alpha$ emission line ($\gtrsim$1000~km/s) compared with their [N \emissiontype{II}] line, suggesting that those sources are Type 1 AGNs.
Two out of the three broad H$\alpha$ line sources are the ones detected by $Chandra$.
At last, we also tried to distinguish Type 2 AGNs (and LINER) from star-forming galaxies using a diagnostic line ratio diagram referred to as "BPT diagram" \citep{BaldwinPhillipsTerlevich81, Veilleux87, Kewley01, Kauffmann03} with [O \emissiontype{III}]$\lambda$5007/H$\beta$ and [N \emissiontype{II}]/H$\alpha$ line ratios. 
We combined our AKARI-FMOS data with the existing spectroscopic data and found two galaxies have all the four required emission lines. 
\citet{Kewley13} have simulated the chemical evolution with four extreme scenarios, (1) Normal interstellar medium (ISM) condition /metal rich AGN, (2) Normal ISM condition/metal poor AGN, (3) Extreme ISM condition/metal rich AGN, and (4) Extreme ISM condition/metal poor AGN (see their Figs. 1 and 2 for more details), and have predicted how the border between star-forming galaxies and AGNs in the BPT diagram shifts at different redshift.
The both galaxies with the four required emission line detections are located in the star-forming galaxy region in the BPT diagram with adopting any scenarios by \citet{Kewley13}.
The other galaxies unfortunately have only their [N \emissiontype{II}]/H$\alpha$ measured.
Recently, \citet{Kartaltepe15} found in the BPT diagrams that many of far-IR selected (U)LIRGs at around $z\sim1$ look like they are dominated by AGN (or at least are composite systems), and that their results are well in agreement with the scenario 1, 3, and 4 of \citet{Kewley13}.
They also reported that some of the BPT-selected AGNs are not detected by their X-ray emission. 
In the scenario 1 and 3 of \citet{Kewley13} models, almost all galaxies with [N \emissiontype{II}]/H$\alpha~<-0.4$ come into the star-formation region regardless of the [O \emissiontype{III}]$\lambda$5007/H$\beta$ ratio.
When the $-0.4<$ [N \emissiontype{II}]/H$\alpha<-0.3$ about half of galaxies fall into the AGN region, and most of galaxies with the ratio of $>-0.3$ come into the AGN region. 
In the scenario 4, even though the [N \emissiontype{II}]/H$\alpha$ is $\sim-0.5$, about half (or more) of galaxies are probably distributed in the AGN area.
Thus, considering the [N \emissiontype{II}]/H$\alpha$ ratios of our sample, 30\% -- 50\% of our sample may be classified as Type 2 AGNs or LINERs depending of ISM conditions and metallicities.

Summarizing the above, we excluded three AGN-affected sources from the following discussions.
In stacked spectra of our sample after this AGN removal, which will be discussed more detail in \S \ref{sec:stackingana}, we cannot find any clear [O \emissiontype{I}]$\lambda$6300 emission line signal.
According to a diagnostic with [O \emissiontype{III}]$\lambda$5007/H$\beta$ and [O \emissiontype{I}]$\lambda$6300/H$\alpha$ line ratio \citep{Kewley06}, objects with [O \emissiontype{I}]$\lambda$6300/H$\alpha$ $<$ -- 0.9 can be classified as star-forming galaxies depending on their [O \emissiontype{III}]$\lambda$5007/H$\beta$.
Using noise levels of the continuum emission of the stacked spectra with a line width of an average [N \emissiontype{II}] and [S \emissiontype{II}] narrow lines, 3$\sigma$ upper limit of the [O \emissiontype{I}]$\lambda$6300/H$\alpha$ line ratios are $\sim -0.8$.
Therefore, even though the no [O \emissiontype{I}]$\lambda$6300 signal in the spectra which supports that our sample is dominated by star-forming galaxies, we need to remember that the our final sample may still include AGN-affected galaxies which result in overestimations of their metallicities.

\section{Measurements}
\subsection{Stellar Mass and Infrared Luminosity}
\label{sec:LIR}

\begin{table}[htbp]
\begin{center}
\caption{{Line fluxes, and physical parameters.\hfil\hfill}}
\label{tb:detail}
\small
\begin{tabular}{lcrcccccc}
\hline
  \multicolumn{1}{c}{Object} &
  \multicolumn{1}{c}{Redshift} &
  \multicolumn{1}{c}{$f_{{\rm H}\alpha}$} &
  \multicolumn{1}{c}{$f_{[N \emissiontype{II}]\lambda 6584}$} &
  \multicolumn{1}{c}{log~($M_{\ast}$/M$_{\Sol}$)} &
  \multicolumn{1}{c}{log~($L_{\rm IR}$/L$_{\Sol}$)} &
  \multicolumn{1}{c}{SFR} &
  \multicolumn{1}{c}{$A_{\rm H\alpha}$} &
  \multicolumn{1}{c}{12+log({\rm O/H})} \\
  \multicolumn{1}{c}{(1)} &
  \multicolumn{1}{c}{(2)} &
  \multicolumn{1}{c}{(3)} &
  \multicolumn{1}{c}{(4)} &
  \multicolumn{1}{c}{(5)} &
  \multicolumn{1}{c}{(6)} &
  \multicolumn{1}{c}{(7)} &
  \multicolumn{1}{c}{(8)} &
  \multicolumn{1}{c}{(9)}\\
  \hline
61007260         & 0.704 &   3.41 $\pm$ 0.43 & $<$   1.16           & 10.40 & 11.72 & 25.73 $\pm$ 7.35 & 2.17        & $<$ 8.63                           \\ 
61010028         & 0.769 &   2.70 $\pm$ 0.30 & $<$   0.60           &   9.96 & 11.63 & 21.54 $\pm$ 6.10 & 1.99        & $<$ 8.53                           \\ 
61010435         & 0.809 &   2.26 $\pm$ 0.33 & 0.69 $\pm$ 0.23 & 10.10 & 10.90 &  6.70 $\pm$ 1.81 & 0.78        &        8.61 $\pm$ 0.08        \\ 
61010515         & 0.922 &   3.09 $\pm$ 0.53 & 1.11 $\pm$ 0.37 & 10.73 & 11.83 & 35.33 $\pm$ 8.45 & 1.90        &        8.65 $\pm$ 0.09         \\ 
61011420         & 0.768 &   9.14 $\pm$ 0.67 & 5.06 $\pm$ 0.71 & 10.74 & 11.78 & 37.40 $\pm$ 6.75 & 1.27        &        8.75 $\pm$ 0.04$^b$ \\ 
61011677         & 0.885 &   5.48 $\pm$ 0.61 & 1.56 $\pm$ 0.20 &   9.80 & 11.99 & 51.38 $\pm$ 12.58 & 1.79        &        8.59 $\pm$ 0.02         \\ 
61012118         & 0.955 &   2.49 $\pm$ 0.41 & 1.25 $\pm$ 0.29 & 10.31 & 11.77 & 30.41 $\pm$ 14.34 & 1.87        &        8.73 $\pm$ 0.05         \\ 
61012132         & 0.956 &   3.39 $\pm$ 0.39 & $<$   0.78           & 10.21 & 11.76 & 31.83 $\pm$ 12.65 & 1.59        & $<$ 8.54                            \\ 
61012133         & 0.959 &   2.42 $\pm$ 0.42 & 0.97 $\pm$ 0.29 & 10.27 & 11.48 & 18.18 $\pm$ 3.80 & 1.33        &        8.67 $\pm$ 0.07         \\ 
61012385$^a$ & 0.784 &   3.03 $\pm$ 0.59 & 1.02 $\pm$ 0.20 & 10.54 & 11.67 &     $\cdots$           &  $\cdots$ & $\cdots$                            \\ 
61013116         & 0.923 &   3.72 $\pm$ 0.54 & 1.38 $\pm$ 0.31 & 10.20 & 11.77 & 32.75 $\pm$ 8.24 & 1.61        &        8.65 $\pm$ 0.06         \\ 
61013936         & 1.029 &   1.50 $\pm$ 0.36 & 1.15 $\pm$ 0.33 & 10.63 & 11.48 & 16.89 $\pm$ 3.57 & 1.59        &        8.83 $\pm$ 0.08$^b$ \\ 
61014553         & 0.923 &   3.91 $\pm$ 0.55 & $<$   1.05           & 10.32 & 11.28 & 15.99 $\pm$ 3.06 & 0.78        & $<$ 8.57                            \\ 
61016374         & 0.910 &   1.99 $\pm$ 0.34 & 0.73 $\pm$ 0.23 &  10.16 & 11.48 & 16.73 $\pm$ 5.13 & 1.60        &        8.65 $\pm$ 0.07         \\ 
61017060         & 0.868 &   2.66 $\pm$ 0.32 & $<$   0.73           & 10.55 & 11.58 & 20.70 $\pm$ 4.78 & 1.64        & $<$ 8.58                            \\ 
61017881         & 0.950 &   2.80 $\pm$ 0.25 & 1.12 $\pm$ 0.21 & 10.97 & 12.11 & 61.56 $\pm$ 18.03 & 2.53        &        8.67 $\pm$ 0.05         \\ 
61018324$^a$ & 1.031 & 21.42 $\pm$ 3.38 & 0.68 $\pm$ 0.14 & 10.29 & 11.52 & $\cdots$\if015.82\fi & $\cdots$ & $\cdots$      \\ 
61019568         & 0.893 &   2.21 $\pm$ 0.33 & $<$   0.57           &   9.88 & 11.54 & 18.93 $\pm$ 8.10 & 1.67        & $<$ 8.56                           \\ 
61020367$^a$ & 0.923 & 10.32 $\pm$ 3.62 &        $\cdots$       & 10.34 & 11.47 &       $\cdots$          & $\cdots$ & $\cdots$                            \\ 
61020444         & 0.732 &   9.39 $\pm$ 1.30 & $<$   2.61           & 10.88 & 11.76 & 35.04 $\pm$ 7.11 & 1.30        & $<$ 8.58                           \\ 
61021272         & 0.824 &   2.38 $\pm$ 0.29 & 1.06 $\pm$ 0.23 & 10.41 & 11.38 & 13.89 $\pm$ 2.70 & 1.47        &        8.70 $\pm$ 0.06        \\ 
61022683         & 1.022 &   3.10 $\pm$ 0.54 & 1.18 $\pm$ 0.35 & 10.40 & 11.65 & 27.16 $\pm$ 8.65 & 1.33        &        8.66 $\pm$ 0.07        \\ 
61022934         & 0.703 &   7.61 $\pm$ 0.93 & 3.70 $\pm$ 0.59 & 10.53 & 11.38 & 18.02 $\pm$ 2.83 & 0.91        &        8.72 $\pm$ 0.04        \\ 
61023221         & 0.703 &   4.09 $\pm$ 0.50 & 1.02 $\pm$ 0.26 &   9.89 & 10.86 &   7.27 $\pm$ 1.11 & 0.60        &        8.56 $\pm$ 0.06        \\ 
61023846         & 0.795 &   2.62 $\pm$ 0.34 & 1.27 $\pm$ 0.29 & 10.68 & 11.39 & 14.09 $\pm$ 2.85 & 1.48        &        8.72 $\pm$ 0.05        \\ 
61024055         & 1.025 & 19.48 $\pm$ 2.34 & 3.12 $\pm$ 0.47 &  9.96 & 11.83 & 79.63 $\pm$ 12.44 & 0.50        &        8.45 $\pm$ 0.02        \\ 
61024136         & 0.927 &   2.00 $\pm$ 0.34 &           $<$   0.66 & 10.05 & 12.00 & 46.58 $\pm$ 12.21 & 2.66        & $<$ 8.63                          \\ 
61024723         & 1.030 &   2.47 $\pm$ 0.50 & 1.67 $\pm$ 0.44 & 10.14 & 11.51 & 20.39 $\pm$ 3.84 & 1.25        &        8.80 $\pm$ 0.07$^b$\\ 
\hline
\end{tabular}
\end{center}
(1): AKARI ID. (2): Redshift derived from the FMOS spectrum. (3) and (4): Fiber aperture corrected H$\alpha$ and [N \emissiontype{II}] emission line fluxes in $10^{-16}$~erg~s$^{-1}$~cm$^{-2}$. If [N \emissiontype{II}] emission is not detected with $S/N>3$, then 3$\sigma$ upper limit flux is summarized. (5) and (6): Stellar mass and IR luminosity calculated by SED fitting with the FMOS $z_s$. (7): SFR estimated using $L_{\rm H\alpha, obs}$ and $L_{\rm IR}$ in M$_{\Sol}$~yr$^{-1}$. (8): H$\alpha$ extinction derived with the ratio of $L_{\rm H\alpha, obs}$ and $L_{\rm IR}$. (9): Metallicity using N2 index (= log~([N \emissiontype{II}]$\lambda$6584/H$\alpha$)).\\
$^a$ Broad H$\alpha$ line sources. The broad H$\alpha$ fluxes are not used for SFRs and metallicity measurements.\\
$^b$ N2 index is larger than $-$0.3.
\end{table}

To measure more reliable stellar mass and infrared luminosity, we re-calculate these parameters with FMOS spectroscopic redshifts and $Herschel$/PACS and SPIRE photometric data.
$Herschel$/PACS (100 and 160~$\mu$m) and SPIRE (250, 350, and 500~$\mu$m) have carried out deep imaging
of the AKARI NEP-Deep field (Pearson et al. in prep.).
We estimate the fluxes in each of the five $Herschel$ bands, and consider the photometric data with $S/N>1.5$ as detected (see Table \ref{tb:detail}). 
Then we re-run the SED fitting for measuring stellar mass ($M_{\ast}$) and IR luminosity ($L_{\rm IR}$) including 22 photometries from $u^{*}$ to 500~$\mu$m using the same parameter setting of $Le~Phare$ code as that we used in \S\ref{Photometric Data in the AKARI NEP-Deep field}.
Actually the re-calculated $M_{\ast}$ and $L_{\rm IR}$ are not much different from the original values.
The average (the standard deviation) of the differences of $M_{\ast}$ and $L_{\rm IR}$ are 0.02~dex (0.16~dex) and 0.06~dex (0.22~dex), respectively.
The derived $M_{\ast}$ and $L_{\rm IR}$ are in ranges of log($M_{\ast}$/M$_{\Sol}$) = 9.52 -- 11.09 ($\langle$log($M_{\ast}$/M$_{\Sol}$)$\rangle$ = 10.34) and log($L_{\rm IR}$/L$_{\Sol}$) = 10.86 -- 12.18 ($\langle$log($L_{\rm IR}$/L$_{\Sol}$)$\rangle$ = 11.58), summarized in Table \ref{tb:detail}.

\subsection{SFR and Dust extinction}
\label{sec:SFRextinction}

We used two ways to estimate SFR for each of our sample.
One is the theoretical relation between SFR and $L_{\rm IR}$ (SFR$_{IR}$) for infrared bright star-bursting galaxies assuming that the dust in the galaxies is heated by stars with ages less than 30~Myr and not taking account of the effect on the dust heating by older stars \cite{Kennicutt98}, which we used in $\S$ \ref{Photometric Data in the AKARI NEP-Deep field}.
The other one is empirical relations between SFR and a linear combination of observed (= attenuated) H$\alpha$ luminosity and $L_{\rm IR}$ \citep{Kennicutt09} for local normal star-forming galaxies ($\log L_{\rm IR}/L_{\odot} < 11.9$) as follows:
\begin{eqnarray}
L_{\rm H\alpha, int} &=& L_{\rm H\alpha, obs} + \alpha L_{\rm IR},
\label{eq:LHaint}\\
{\rm SFR}_{H\alpha, IR} &=& (7.9/1.7)\times 10^{-42}~L_{\rm H\alpha, int}~({\rm erg~s^{-1}}),
\label{eq:SFRHaIR}
\end{eqnarray}
where a scaling coefficient $\alpha$ is $(2.4 \pm 0.6) \times 10^{-3}$ based on the best agreement with the luminosities corrected from the Balmer decrements, and the SFR is corrected to the \citet{Chabrier03} IMF by dividing by 1.7.
We used aperture corrected H$\alpha$ luminosity and $L_{IR}$ derived above for Eq. (\ref{eq:LHaint}).
Average SFR$_{H\alpha, IR}$ and SFR$_{IR}$ are 25.2~M$_{\Sol}~{\rm yr}^{-1}$ and 47.7~M$_{\Sol}~{\rm yr}^{-1}$, respectively.
When we compare SFR$_{H\alpha, IR}$ with SFR$_{IR}$ for individual sources, we found systematic difference by $\sim 2.2$.
\citet{Kennicutt09} already mentioned about the inconsistency, and suggested that we use a different equation depending on a nature of our science targets.
Since most of the galaxies in our sample are classified as LIRGs, in this paper we basically use SFR$_{H\alpha, IR}$, and use SFR$_{IR}$ supplementary.

\begin{figure}[htbp]
\begin{center}
  \includegraphics[width=100mm]{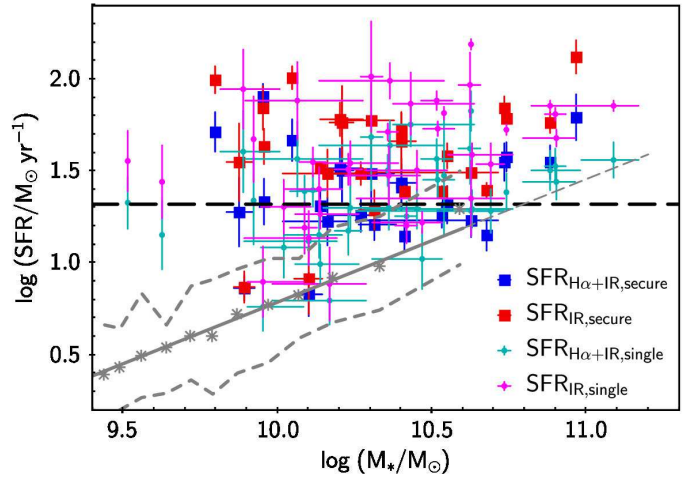}
\caption{The $M_{\ast}$--SFR relation derived of AKARI--FMOS sample (color points). Bigger indices are secure H$\alpha$ detected sources and smaller ones represent single H$\alpha$ detected sources. Bluer color-coded indices mean SFRs measured with $L_{\rm H\alpha, obs}$ and $L_{\rm IR}$, while redder color-coded ones show SFRs calculated with only $L_{\rm IR}$. MS galaxies at $z\sim0.78$ \citep{Zahid12} are over plotted as a comparison with gray symbols (the solid and dashed lines represent the best fit and the 68\% confidence interval of their galaxies). The black long dashed line represents the predicted detection limit of our FMOS $J$-long observation with 1 hour integration for galaxies at $z\sim0.88$.}
  \label{fig:MS}
\end{center}
  \end{figure}

Figure \ref{fig:MS} shows the SFRs of the AKARI-FMOS objects as a function of their stellar mass.
A tight relationship between stellar mass and SFR for DEEP2 star-forming galaxies (so called main sequence galaxies; MS galaxies) at $z\sim0.78$ \citep{Zahid12} is overplotted.
Apparently the SFR of our AKARI-FMOS sources show no clear correlation with stellar mass.
Also, no significant difference of SFRs between secure H$\alpha$ detected sources and single H$\alpha$ detected sources are seen.

We calculate an average detection limit of our FMOS observation. 
The average exposure time of one target is 60 minutes, corresponding detection limit of line flux is $0.4 \times 10^{-16}$~erg~s$^{-1}$~cm$^{-2}$.
Since the average redshift of our sample is $z\sim0.88$, an average $L_{\rm H\alpha, obs}$ is $1.33 \times10^{41}$~erg~s$^{-1}$.
Therefore, the corresponding $S/N=5$ SFR detection limit for our AKARI-FMOS sample at $z\sim0.88$ is log (SFR) = 1.32 M$_{\Sol}$~yr$^{-1}$ (black dashed line in Figure \ref{fig:MS}),  taking account of the average $L_{\rm IR}$ calculated in $\S$\ref{sec:LIR} of $4.78\times 10^{11}$ L$_{\Sol}$.
It is obvious that the average detection limit of our observation does not reach to detect the SFR of MS galaxies at $z\sim0.78$ across most of the stellar mass range, especially at the lower stellar masses.
Therefore, we conclude that our flat $M_{\ast}$--SFR relation is due to the detection limit of our FMOS observations.

The ratio of $L_{\rm IR}$ and $L_{\rm H\alpha, obs}$ can be used to trace extinction, $A_{\rm H\alpha}=2.5\times {\rm log}\left(1+\alpha \frac{L_{\rm IR}}{L_{\rm H\alpha, obs}}\right)~[{\rm mag}]$, where the $\alpha$ is the same scaling coefficient as that for Eq. (\ref{eq:LHaint}).
The calculated extinctions of our sample are summarized in Table \ref{tb:detail}.
The $A_{\rm H\alpha}$ range of our sample is $0.5<A_{\rm H\alpha}<2.7$, and average (median) is $A_{\rm H\alpha}$  = 1.50 (1.59).
The extinction range estimated from the SED fitting is $0.9<A_{\rm H\alpha}<2.4$, and average (median) is $A_{\rm H\alpha}$  = 1.29 (1.20).
Between the $A_{\rm H\alpha}$ estimated using H$\alpha$ emission line and using continuum emission (SED fitting), the range is not much different but the average (median) value from the SED fitting is smaller by about 0.2 -- 0.3 magnitude.
For sources with $A_{\rm H\alpha} < 1.5$ corresponding to the $E(B-V)$ of 0.5, the average of the difference is 0.006~mag, which is well match with each other, while the value is 0.3~mag for sources with $A_{\rm H\alpha} >1.5$.
This could reflect the difficulty of estimating the extinction of line emitting region from the extinction of continuum emitting region.
As a reference, typical $A_{\rm H\alpha}$ using an equation proposed by \citet{GarnBest10} for HiZELS H$\alpha$ selected sample at $z\sim0.84$ is 1.67, 
similar to our result.
\citet{Buat15} have plotted our result converted to $A_{\rm UV}$ in Figure 10 of their paper and have compared the extinction with their result along with previous publications. 
The extinction of our AKARI-FMOS sample is almost a typical value of IR selected samples but larger than a global dust extinction at the same redshift range \citep{Burgarella13}.

\subsection{Metallicity}
\label{sec:metallicity}

There are some widely used methods for determining the abundance of metallicity from both theoretical approaches based on photoionization models and empirical approaches based on temperature-sensitive optical emission line ratios.
In this work, we determine metallicities using the N2 index (= log~([N \emissiontype{II}]$\lambda$6584/H$\alpha$)) calibrated by \citet{PettiniPagel04}.
From a fitting of observed relationship between N2 value and metallicity for a sample of 137 H \emissiontype{II} regions, they found 
\begin{eqnarray}
12+{\rm log} ({\rm O/H})=8.90+0.57\times {\rm N2}, 
\label{eq:N2indicator}
\end{eqnarray}
with a 1$\sigma$ dispersion of 0.18~dex.
An advantage of this method is the required lines for the N2 index are close in wavelength. 
Thus, these lines can be easily observed simultaneously, and little extinction correction and little aperture correction are needed.
The metallicity calculation is valid when N2 index is smaller than $-$0.3 because N2 index saturates when nitrogen becomes the dominant coolant.
Although there are three sources in our sample whose N2 values reach the limit, we do not exclude them from following discussion because their N2 values could be smaller than $-$0.3 within the uncertainties. 
Different metallicity calibrations with different strong emission line ratios yield systematically different abundances.
Using $\sim$28,000 galaxies from the SDSS DR4, \citet{KewleyEllison08} derived quartic functions for converting metallicities between various diagnostics.
We use these formulae to convert metallicities determined from various diagnostics used in previous studies to the N2 diagnostic for comparing our metallicities.
In our work, the errors are estimated from only fitting errors of continuum and emission lines, and we do not include any possible intrinsic errors such as metallicity calibration, systematic differences between various metallicity indicators, fiber position accuracy, aperture correction factors.
Hence, the errors can be recognized as a lower limit.

\subsection{Stacking Analysis}
\label{sec:stackingana}

\begin{figure}[htbp]
  \begin{center}
    \includegraphics[width=70mm,angle=-90]{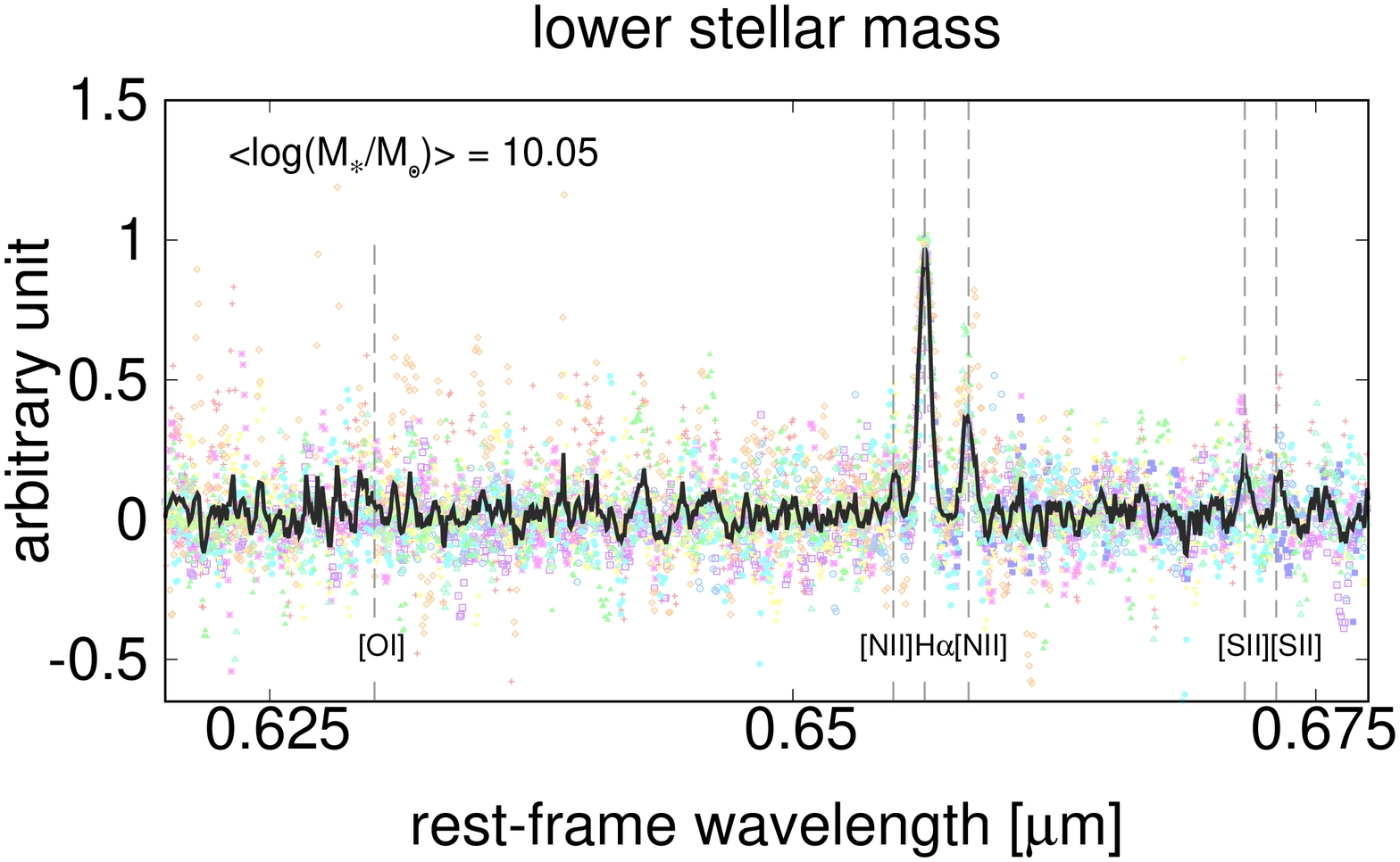}
    \includegraphics[width=70mm,angle=-90]{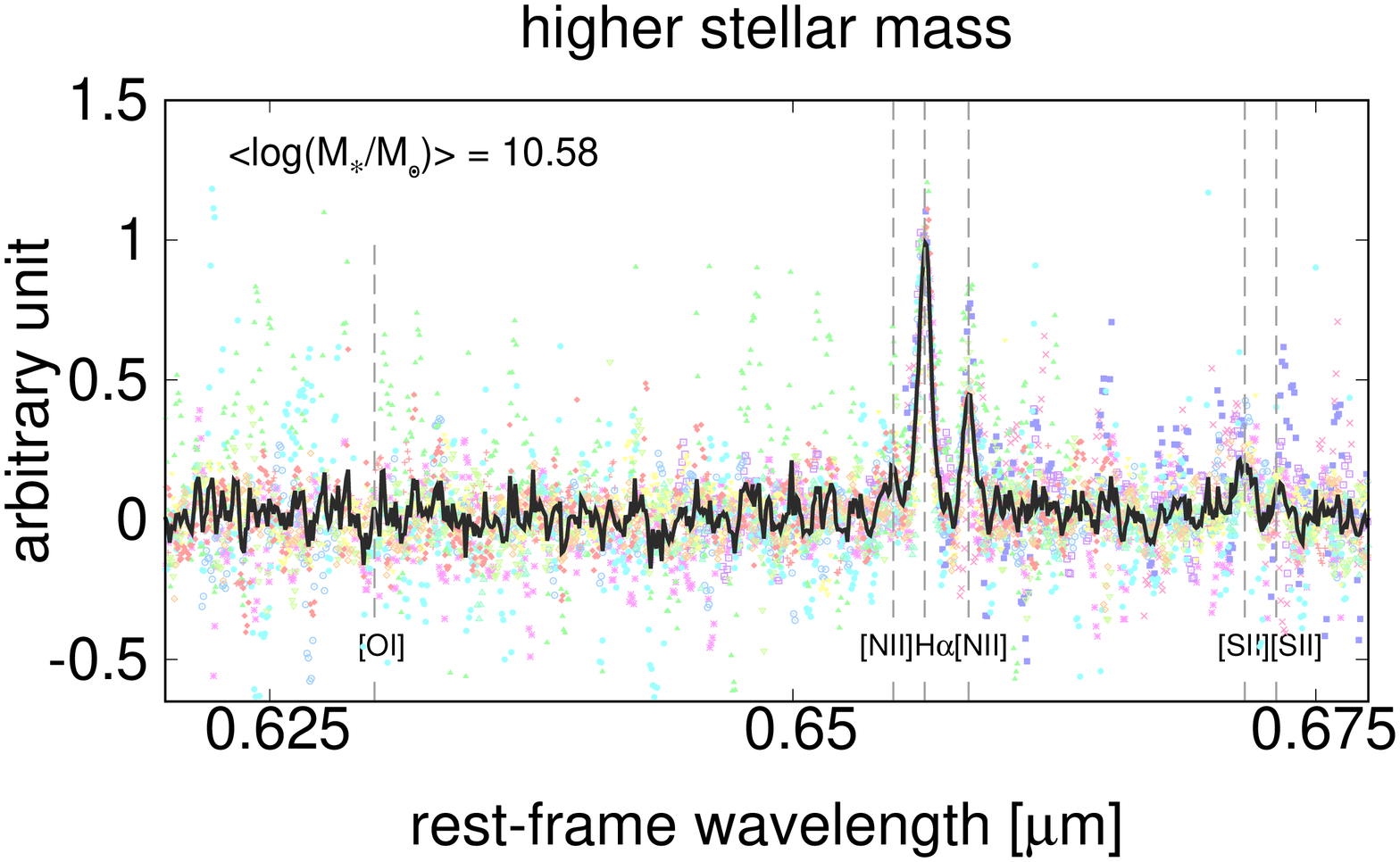}
    \caption{H$\alpha$-peak normalized staking spectra. Lower stellar mass subsample of 12 galaxies and higher stellar mass subsample of 13 galaxies are plotted in color points in upper and lower panels, respectively. An averaged spectrum of each subsample is overlaid in each panel (thick line). The average stellar mass value for each subsample is shown in the top left of each panel.}
    \label{fig:StackSpectra}
  \end{center}
\end{figure}

\begin{figure}[htbp]
  \begin{center}
    \includegraphics[width=70mm,angle=-90]{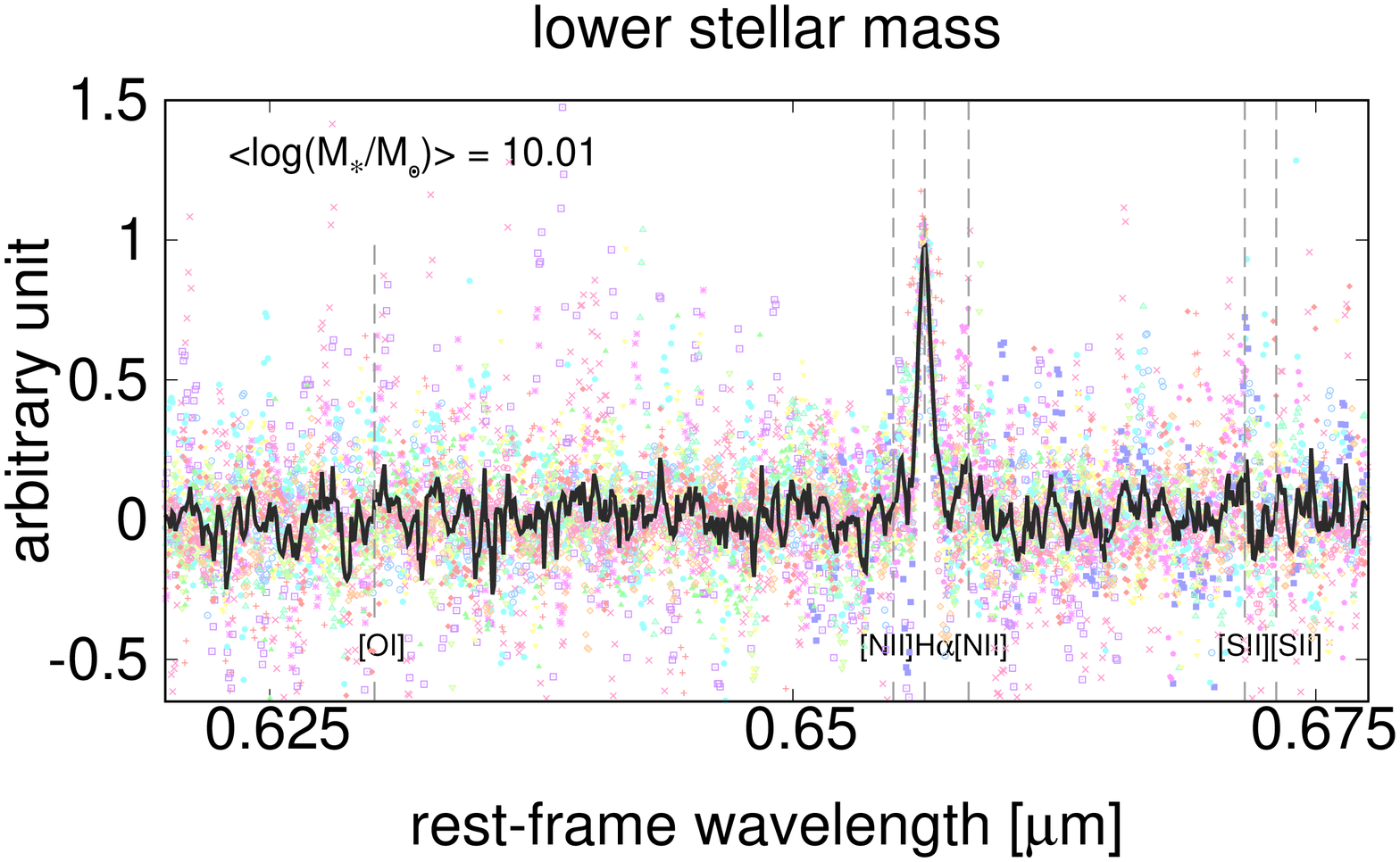}
    \includegraphics[width=70mm,angle=-90]{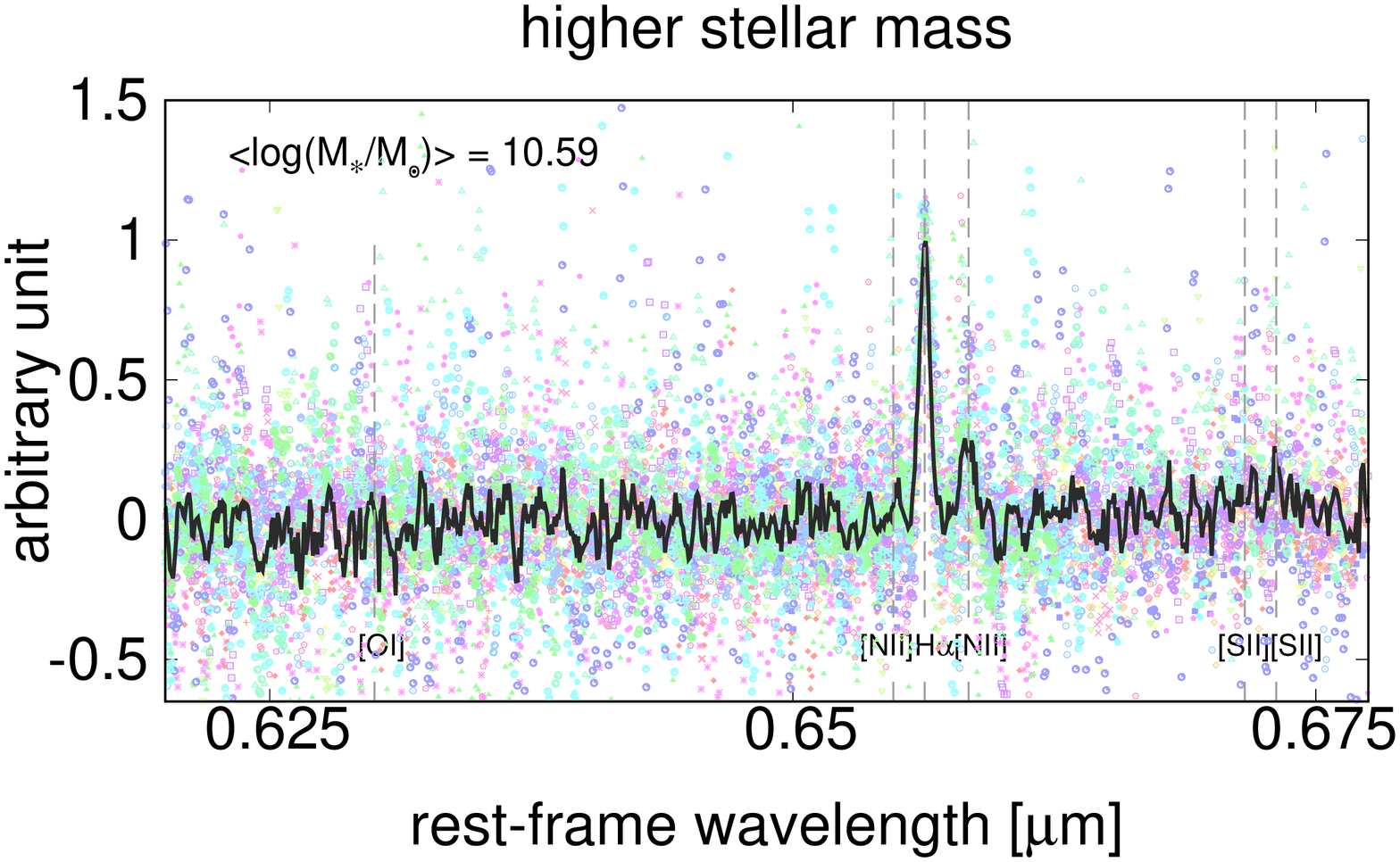}
    \caption{The same figure as Figure \ref{fig:StackSpectra}, but for single H$\alpha$ emission line detected sources. Top and bottom figure are for low $M_{\ast}$ subsample (15 sources) and high $M_{\ast}$ subsample (21 sources) separated at $M_{\ast}=10.30$ as the same for the secure H$\alpha$ detected sources.}
    \label{fig:StackSpectrasingle}
  \end{center}
\end{figure}

One third (= 8/25) of our secure H$\alpha$ detected sources have a 3$\sigma$ upper limit of metallicities due to the upper limits on
their [N \emissiontype{II}]. 
If we focus on the metallicity-derived sources, our results must be biased toward high metallicity.
In order to measure a typical metallicity for our sample at $z\sim0.88$, we split the secure H$\alpha$ detected sample into two stellar mass bins, stack the individual spectra, and measure the metallicity at the stellar mass in each bin.
The mass ranges and average values are $9.80<$ log~$M_{\ast}$/M$_{\Sol}<10.27$ and $\langle {\rm log (M}_{\ast}/M_{\Sol})\rangle = 10.05$ for 12 low mass galaxies and $10.31<$ log$M_{\ast}$/M$_{\Sol}<11.97$ and $\langle {\rm log (M}_{\ast}/M_{\Sol})\rangle = 10.58$ for 13 high mass galaxies.
First, we subtract a continuum emission derived in \S\ref{sec:LineFitting} from each observed spectrum, replace observed fluxes at destroyed wavelengths by the OH masks to zero fluxes, normalize the spectra by the peak flux of H$\alpha$ emission, and de-redshift the spectra to the rest-frame wavelength with their redshifts. 
Next, we average the spectra without any weight.
The stacked-spectrum in each bin is exhibited in Figure \ref{fig:StackSpectra}.
Finally, we measure continuum levels of the stacked-spectra using fluxes at 0.610 -- 0.625~$\mu$m and 0.661 -- 0.670~$\mu$m, and fit Gaussian functions to the H$\alpha$ and [N \emissiontype{II}] emission. 
The results are summarized in Table \ref{tb:stack}.
We also made median-stack spectra for both stellar mass bins, and we found almost no difference between the line ratios from averaged and median stacked spectra.

Moreover, we obtain the average stacked spectra for the 36 single emission line detected sources with the assumption that the emission line is H$\alpha$, although we cannot know what line they truly are.
For the stacking analysis, we divide the 36 objects into two subsamples depending on their stellar masses, and use the same manner as we used for the secure H$\alpha$ detected sources.
We adopt the stellar mass of $M_{\ast}$/M$_{\odot}$ = 10.30 to separate the sample to high and low stellar mass bins, which is the same for the secure H$\alpha$ detected sources.
The low mass bin has 15 sources and high mass bin has 21 sources. 
From Figure \ref{fig:StackSpectrasingle}, [N \emissiontype{II}]$\lambda$6584 can be seen from the both stacked spectra with $S/N>3$ and almost no clear signals are found at wavelengths of [S \emissiontype{II}] doublet.
It may be due to the following reason: Many of the detected single emission lines are actually H$\alpha$ but some are not.
Because [S \emissiontype{II}] lines are generally weaker than [N \emissiontype{II}] line, the [N \emissiontype{II}] line survives while the [S \emissiontype{II}] is buried in noise.
Thus, the line fluxes measured from the stacked spectra are more likely lower than the real value.
Furthermore, we calculate the stacked spectra using all 61 H$\alpha$ detected sources. 
The results are also shown in Table \ref{tb:stack}.

\begin{table}[htbp]
\begin{center}
\caption{Emission line ratios and metallicity from stacked spectra.\hfil\hfill}
\label{tb:stack}
\small
\begin{tabular}{ccccccc}
\hline
  \multicolumn{1}{c}{log~($M_{\ast}$/M$_{\Sol}$)} &
  \multicolumn{1}{c}{SFR$_{H\alpha, IR}$ (SFR$_{IR}$)~[M$_{\Sol}$~yr$^{-1}$]} &
  \multicolumn{1}{c}{N2} &
  \multicolumn{1}{c}{N2S2} &
  \multicolumn{1}{c}{[S \emissiontype{II}]$\lambda$ 6716/$\lambda$ 6731} &
  \multicolumn{1}{c}{log(N/O)}& 
  \multicolumn{1}{c}{12+log(O/H)}\\
  \hline
    \multicolumn{7}{c}{secure H$\alpha$ detected sources} \\
\hline
10.05$^{+0.22}_{-0.25}$ & 29.33 (47.95) $\pm$ 8.43 (13.50) & $-$0.46 $\pm$ 0.09 &  0.10 $\pm$ 0.16 & 1.46 $\pm$ 0.98 & $-$0.74 $\pm$ 0.20 & 8.64 $\pm$ 0.05\\
10.58$^{+0.39}_{-0.28}$ & 27.09 (49.27) $\pm$ 8.30 (13.42) & $-$0.38 $\pm$ 0.07 &  0.17 $\pm$ 0.14 & 3.16 $\pm$ 2.76 & $-$0.64 $\pm$ 0.18 & 8.68 $\pm$ 0.04\\
\hline
\multicolumn{7}{c}{single H$\alpha$ detected sources} \\
\hline
10.01$^{+0.22}_{-0.50}$ & 18.55 (32.28) $\pm$ 8.59 (17.44) & $-$0.66 $\pm$ 0.22 &  $\cdots$ & $\cdots$ & $\cdots$ & 8.52 $\pm$ 0.12\\
10.59$^{+0.50}_{-0.29}$ & 31.46 (60.03) $\pm$ 12.21 (20.08) & $-$0.45 $\pm$ 0.15 &  0.04 $\pm$ 0.25 & 0.63 $\pm$ 0.63 & $-$0.81 $\pm$ 0.31 & 8.64 $\pm$ 0.08\\
\hline
\multicolumn{7}{c}{all sources} \\
\hline
10.03$^{+0.24}_{-0.51}$ & 23.34 (39.24) $\pm$ 8.52 (15.81) & $-$0.60 $\pm$ 0.10 &  0.15 $\pm$ 0.21 & 2.05 $\pm$ 2.03  & $-$0.67 $\pm$ 0.26 & 8.56 $\pm$ 0.06\\
10.59$^{+0.50}_{-0.28}$ & 29.79 (55.91) $\pm$ 10.88 (17.83) & $-$0.42 $\pm$ 0.09 &  0.10 $\pm$ 0.16 & 1.16 $\pm$ 0.75 & $-$0.73 $\pm$ 0.21 & 8.66 $\pm$ 0.05\\
\hline
\end{tabular}
\end{center}
\end{table}

\subsection{Electron density}
\label{sec:electrondensity}

The electron density $n_e$ can be measured from the flux ratio of the S$^+$ emission lines, [S \emissiontype{II}]$\lambda$6716/[S \emissiontype{II}$]\lambda$6731 (e.g. \cite{Osterbrock89}).
Some of our observed spectra include the [S \emissiontype{II}]$\lambda \lambda$6716,6731 doublet, and those lines are weakly seen in the some of the stacked spectra.
We fit the  [S \emissiontype{II}] emission lines with two Gaussian functions, and the measured ratios are summarized in Table \ref{tb:stack}.
From both of the stacked spectra for the secure H$\alpha$ detected sources, [S \emissiontype{II}]$\lambda$6716 is detected with $S/N>3$, while [S \emissiontype{II}]$\lambda$6731 is with $S/N<2$.
We obtained the electron densities using an electron temperature of $T_e=10^4$ K.
This electron temperature is commonly assumed for typical H \emissiontype{II} region, and the dependence of the electron density measured from the [S \emissiontype{II}]$\lambda$6716/[S \emissiontype{II}]$\lambda$6731 line ratio on electron temperature is not severe for the electron temperature \citep{Copetti00}. 
The obtained line ratios from secure H$\alpha$ detected sources are 1.46$\pm$0.98 and 3.16$\pm$2.76, where is corresponding to the electron densities of $\sim10^{1-2}$ cm$^{-3}$ and $<10$ cm$^{-3}$ for low and high mass bins, respectively.
We will discuss about the electron density in $\S$ \ref{sec:compElecDens}.

\section{Discussion}
\subsection{[N \emissiontype{II}]/H$\alpha$ dependences on physical properties}
\label{sec:NIIHadepend}

In this paper, we estimate metallicity (= gas-phase oxygen abundance) using the [N \emissiontype{II}]/H$\alpha$ line ratio.
It is not clear that the relation between the [N \emissiontype{II}]/H$\alpha$ ratio and metallicity--which is calibrated with optical-selected local galaxies--can hold for our sample of active star-forming IR galaxies at high redshift.
In this section, we discuss whether the relation of the [N \emissiontype{II}]/H$\alpha$ ratio and metallicity (i.e., Eq. (\ref{eq:N2indicator})) can hold for our sample. 

\subsubsection{Electron density comparison with other galaxies}
\label{sec:compElecDens}
\citet{Kewley13} found that higher gas density (pressure) increases the [N \emissiontype{II}]/H$\alpha$ line ratio (see also \cite{Dopita16, Kashino17}).
If the density of ionized region of our sample is higher or lower than the local normal star-forming galaxies, the metallicity we measured can be over or under estimated.

Left panel of Figure \ref{fig:NIIenrich} shows the electron density we measured in $\S$ \ref{sec:electrondensity} against their stellar mass.
The contours overplotted in the figure is a distribution of local normal star-forming galaxies from SDSS DR12 survey \citep{Alam15}, showing the typical [S \emissiontype{II}] line ratio is $\sim$ 1.3 -- 1.4, corresponding to $n_e$ $\sim$10--100 cm$^{-3}$.
Within the error bars, our results and typical electron density for the local normal star-forming galaxies are matched at a given stellar mass.  
\citet{Yabe15} estimated that the electron density of $\sim 4000$ star-forming galaxies at $z\sim 1.4$ by the same line ratio as our work and found the range of the electron density is $n_e$ = 10--500 cm$^{-3}$, and the median is 28 cm$^{-3}$, which is also consistent with that of our sample.
On the other hand, some studies of the electron density found higher ($\sim 10^{3}$ cm$^{-3}$) electron density for star-forming galaxies at higher redshift (e.g., \cite{Shimakawa15} at $z\sim 2.5$, \cite{Onodera16} at $z\sim 3.3$), which is higher than our results.
There are also some previous studies about the electron densities for local (U)LIRGs.
\citet{Armus04} investigated three local ULIRGs with AGN at their center regions and found their electron densities are $10^{2-3}$ cm$^{-3}$.
\citet{Farrah07} derived the electron density from 53 local ULIRGs using [Ne \emissiontype{V}] emission lines at 14.32 and 24.42 $\mu$m, which is only produced by an AGN due to the high ionization potential.
They found that the electron density is $< 10^{4}$ cm$^{-3}$ which is consistent with local low luminosity AGNs \citep{Sturm02}.
Moreover, Xu et al. (2014, 2015) revealed the electron density of $10^{4}$ cm$^{-3}$ at compact regions at a center of local ULIRGs with hosting an AGN, estimated by spectral line energy distributions of the total CO emission line measured from ALMA high spatial resolution observation.
Recently, the electron density for 122 local (U)LIRGs are also investigated by \citet{Zhao16}, which most of their sample is not effected by an AGN.
They used [N \emissiontype{II}]122$\mu$m/[N \emissiontype{II}]205$\mu$m emission line ratio by $Herschel$ and found the median value of the electron density is 22 cm$^{-3}$.
Therefore, the properties of the electron density of our infrared galaxies at $z\sim 0.88$ is similar to star-forming galaxies at $z < 1.5$ universe, and to star-formation dominant local (U)LIRGs, but smaller that of star-forming galaxies at high-redshift and AGN affected ULIRGs.
We should emphasize that better (higher $S/N$) spectra for the [S \emissiontype{II}] emission lines are needed to study the electron density for high-$z$ IR galaxies.

\subsubsection{Nitrogen to Oxygen abundance ratio}
\label{sec:NOabundance}
Some previous studies cautioned that some physical conditions of galaxies affect the relation of nitrogen-to-oxygen (N/O) abundance ratio at a given O/H.
For example, time delay in releasing nitrogen and oxygen into the ISM makes the N/O ratio small \citep{Olofsson95}.
Oxygen is produced predominantly in massive stars and in and expelled by core-collapse supernovae on a short time scale of $10^{6-7}$~yrs, 
while nitrogen is mainly formed in intermediate-mass stars and returns into the interstellar medium through asymptotic giant branch (AGB) stars with longer time scale of $10^{8-9}$~yrs \citep{KarakasLattanzio14}. 
On the other hand, an enhanced population of Wolf--Rayet (WR) stars, which release large amount of nitrogen, temporarily boost the N/O ratios \citep{Brinchmann08b, Masters14, Shapley15}.

\begin{figure}[htbp]
  \begin{center}
    \includegraphics[width=70mm, height=70mm]{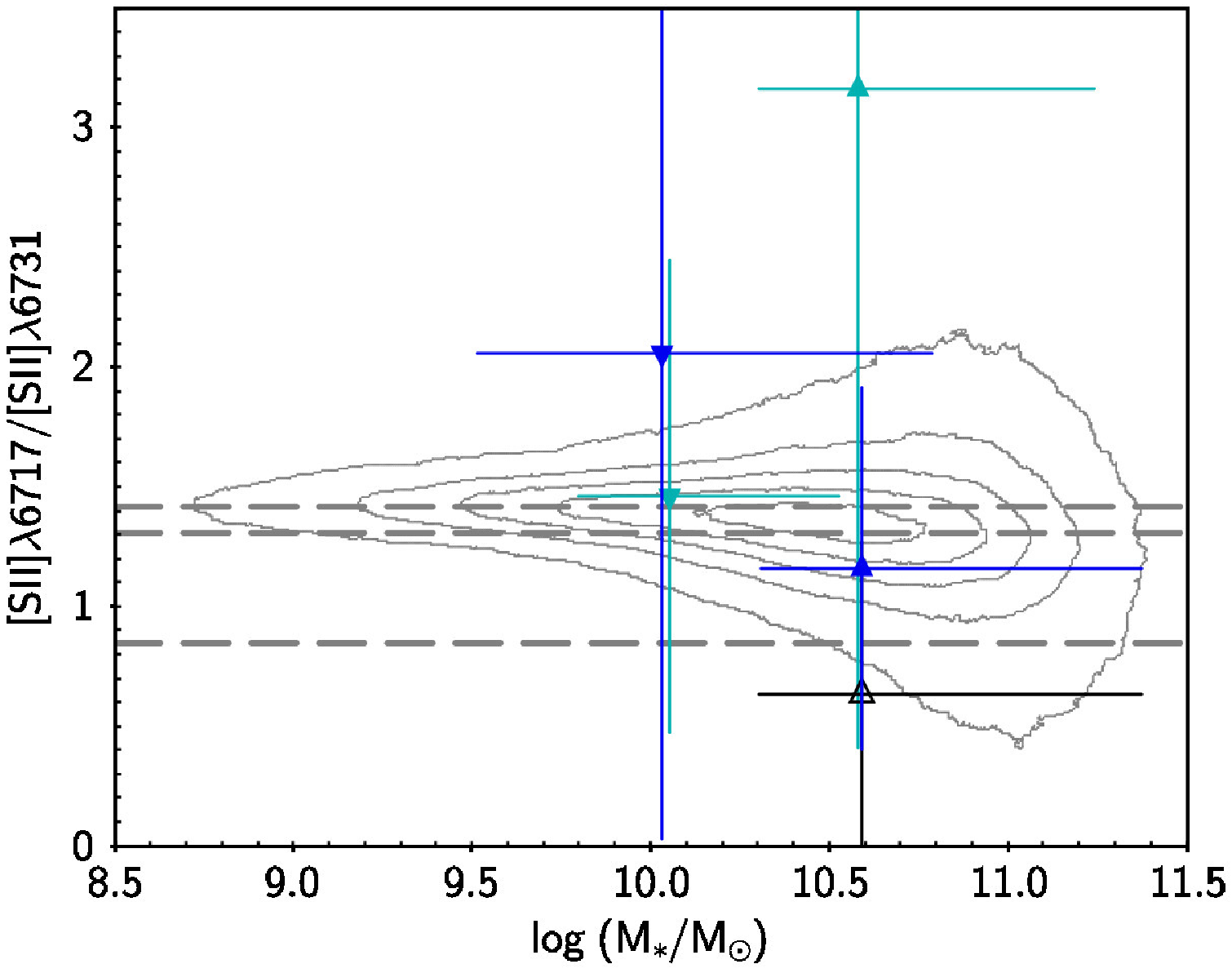}
    \includegraphics[width=70mm, height=70mm]{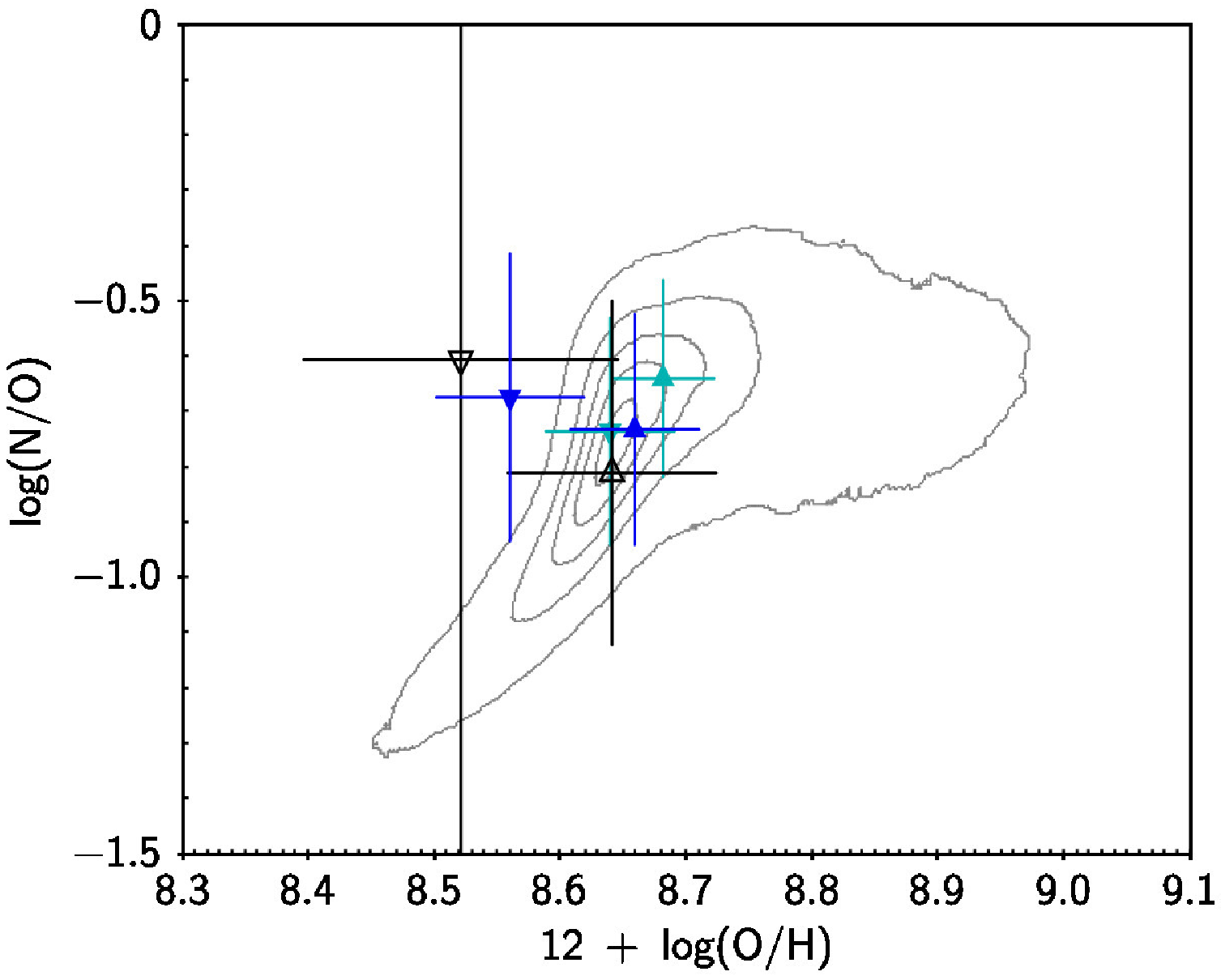}
    \caption{(Left): The [S \emissiontype{II}]$\lambda$6716/[S \emissiontype{II}]$\lambda$6731 line ratio against the stellar mass. The three horizontal dashed lines represent the corresponding electron densities, 10, 100, and 1000 [cm$^{-3}$] from the top, respectively, assuming the electron temperature of 10000K. Upward and downward blue triangles represent stacking results for high and low stellar mass bins, respectively. The results from stacked spectra of secure H$\alpha$ detected sources, single H$\alpha$ detected sources, and all emission detected sources are shown as filled light blue, open black and filled dark blue indices, respectively. The contours show the distribution of the local SDSS galaxies.
    (Right): The N/O ratio as a function of metallicity from stacking spectra. The symbols and contours are the same as the left panel.}
    \label{fig:NIIenrich}
  \end{center}
\end{figure}

To investigate whether the N/O ratio of our IR star-forming (star-bursting) galaxies at $z\sim0.88$ is higher, lower, or consistent with local galaxies, we examine the N/O ratio of our sample using the [S \emissiontype{II}] to [N \emissiontype{II}] line ratio, namely N2S2, measured from our stacking spectra.
The N2S2 index, defined as 
\begin{eqnarray}
{\rm N2S2=log \left\{ \frac{I([N \emissiontype{II}]\lambda6584)}{I([S \emissiontype{II}]\lambda\lambda6716,6731)} \right\},}
\end{eqnarray}
has been proposed as a parameter to estimate the N/O ratio by \citet{Perez-Montero09}.
N2S2 correlates with N/O well because sulphur and oxygen are both $\alpha$ elements, whose origin is primary, while the origin of nitrogen can be both primary and secondary.
They investigated a relation between the N2S2 index and the N/O with local galaxies, and suggested calibration of:
\begin{eqnarray}
{\rm log(N/O)=1.26\times N2S2 - 0.86}
\end{eqnarray}
from the least square linear fit with a standard deviation of 0.31~dex.
The N/O ratios of the subsamples are shown against their metallicities in right panel of Figure \ref{fig:NIIenrich}.
We compare our results with the local SDSS galaxies, and find that the N/O ratios of our high mass subsample agree well with the local 
values a given metallicity, whereas the our low mass subsample shows barely the same level or relatively higher log(N/O), which is seen for high-$z$ galaxies \citep{Masters14, Yabe15}.
Therefore, we conclude that the relation of the [N \emissiontype{II}]/H$\alpha$ ratio and metallicity (i.e., Eq. (\ref{eq:N2indicator})) could hold for our sample or the metallicity might be overestimated especially less massive galaxies.
We should here mention that the N2S2 values and [N \emissiontype{II}]/H$\alpha$ ratio can also be varied by an ionization parameter which can be changed by a slope of IMF or strength of star formation \citep{Kewley13, Dopita13, Dopita16}.
Since unfortunately our spectra do not cover important emission lines useful to estimate the ionization parameters such as [O \emissiontype{II}] and [O \emissiontype{III}], we cannot discuss about the ionization parameter for our sample in this work. 
Thus, more statistical studies at $z\sim1$ and/or direct observation of oxygen emission line would be important for further understanding.

\subsection{The Mass--Metallicity relation of IR galaxies at $z\sim0.88$}

\begin{figure}[htbp]
  \begin{center}
    \includegraphics[width=120mm, angle=-90]{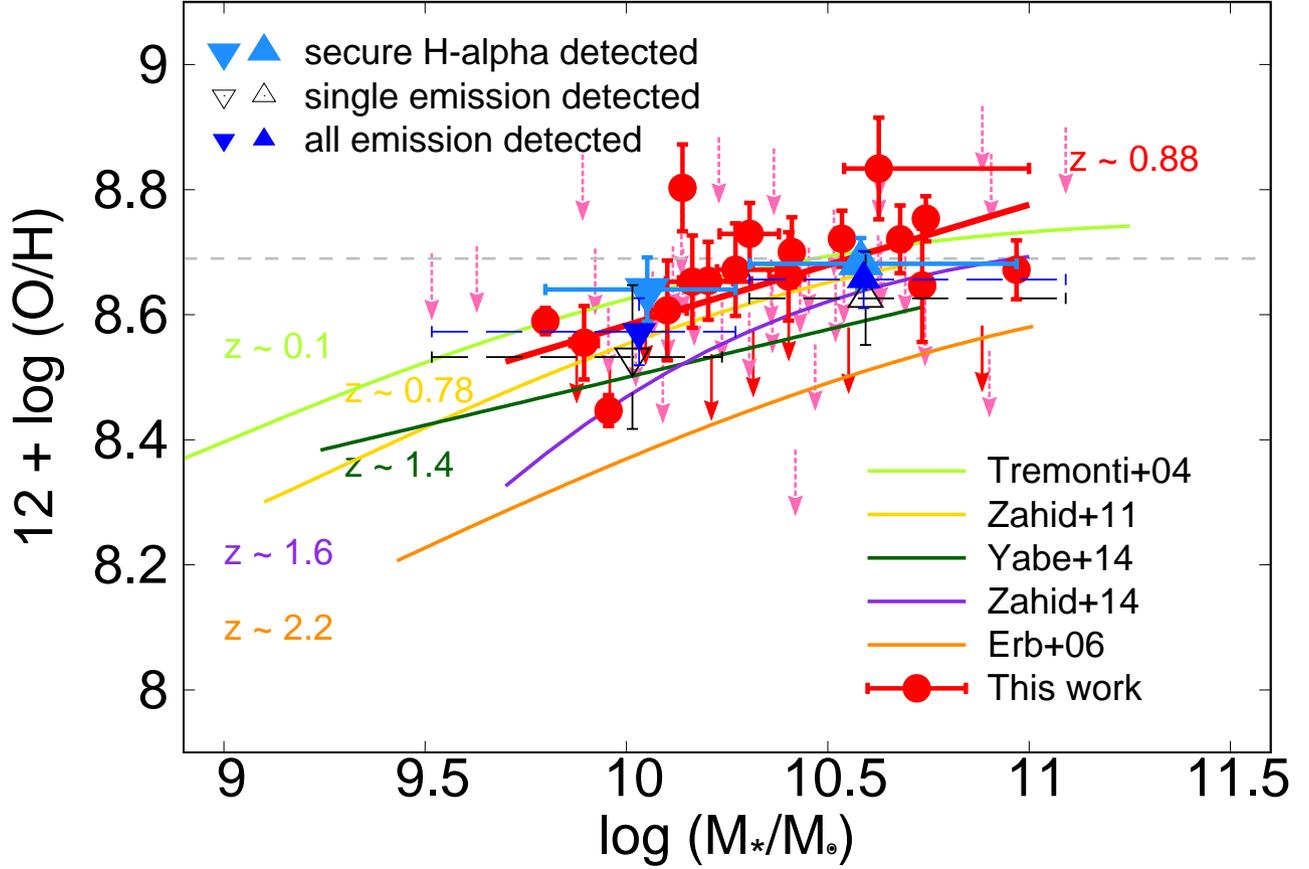}
    \caption{The mass--metallicity relation for our IR star-forming galaxies. The secure H$\alpha$ detected sourced with metallicity measurement individually are plotted in red points, while with 3$\sigma$ limit metallicity measurement are in red arrows.
    The error bars of the metallicities are determined from H$\alpha$ and [N \emissiontype{II}] fitting errors with continuum error, and the 0.18~dex intrinsic uncertainty of the metallicity calibration is not included. 
    The best fit mass--metallicity relation for individually metallicity measured sources is shown with the red thick solid line.
    The pink arrows represent the 3 $\sigma$ upper limit of metallicities for single H$\alpha$ detected sources.
    The metallicity measured from the stacked spectra for low mass and high mass bins are shown with upward and downward triangles, respectively.
    The filled light blue, the opened black, and filled dark blue indices are the same as Figure \ref{fig:NIIenrich}.
    Thin colored lines are the mass--metallicity relations at various redshifts derived from literature; light green:  at $z\sim0.1$ (SDSS; \cite{Tremonti04}), yellow: $z\sim0.78$ (DEEP2; \cite{Zahid11}), dark green: $z\sim1.4$ (SXDS/UDS, $K_{\rm s}$-selection; \cite{Yabe14}), purple: $z\sim1.6$ (COSMOS; \cite{Zahid14}), and orange: $z\sim2.2$ (UV-selection; \cite{Erb06}), respectively. The gray horizontal dashed line indicates solar metallicity (12+log (O/H) = 8.69; \cite{Asplund09}).}
    \label{fig:MZrelation}
  \end{center}
\end{figure}

In Figure \ref{fig:MZrelation}, we plot metallicity as a function of stellar mass for individual galaxies in our sample.
The metallicity derived sources show a clear trend between their stellar masses and metallicities.
From fitting of a mass--metallicity relation for our sample using a linear function with $\chi^{2}$ minimization, the result can be expressed as 
\begin{eqnarray}
12+{\rm log(O/H)} = (7.06\pm0.49)+(0.15\pm0.05)\times{\rm log}(M_{\ast}/{\rm M_{\Sol}}),
\label{eq:MZrelationMoustakas}
\end{eqnarray}
with 0.1~dex dispersion.
We compare our result at $z\sim0.88$ IR galaxies to published observations of galaxies selected at shorter wavelengths (UV to near-IR) 
at various redshifts; at $z\sim0.1$ \citep{Tremonti04}, at $z\sim0.78$ \citep{Zahid11}, at $z\sim1.4$ \citep{Yabe14}, at $z\sim1.6$ \citep{Zahid14}, and at $z\sim2.2$ \citep{Erb06}, converted to the N2 method and \citet{Chabrier03} IMF if needed.
Our results of the secure H$\alpha$ detected sources agree well with the result for normal star-forming galaxies in the local Universe derived by \citet{Tremonti04}.
On the other hand, compared with the mass--metallicity relation at $z\sim0.78$ using DEEP2 survey sample \citep{Zahid11} which is the closest to the redshift of this work among the other data in the literature, the secure H$\alpha$ detected sources seem to have slightly higher metallicity at a given stellar mass.

To investigate the typical metallicity for IR galaxies of our sample, we also plot results of the stacking analysis in Figure \ref{fig:MZrelation}.
We find that the metallicity of high stellar mass galaxies is higher than that of less massive galaxies.
This is consistent with the mass--metallicity trend for our individually measured galaxies.
We also find that the result of the stacking analysis is consistent with the mass--metallicity relations at $z\sim0.1$ and at $z\sim0.78$ within the error bars.
These comparisons suggest that average metallicities of our IR galaxies at $z\sim0.88$ are not significantly different from optically selected MS galaxies at similar redshift. 
Galaxies whose metallicity is high enough to be measured individually are already chemically enriched to the level of local galaxies.

\label{sec:MZrelation}

\subsection{SFR dependence}
\label{sec:SFRdependent}

\begin{figure}[htbp]
  \begin{center}
    \includegraphics[width=90mm]{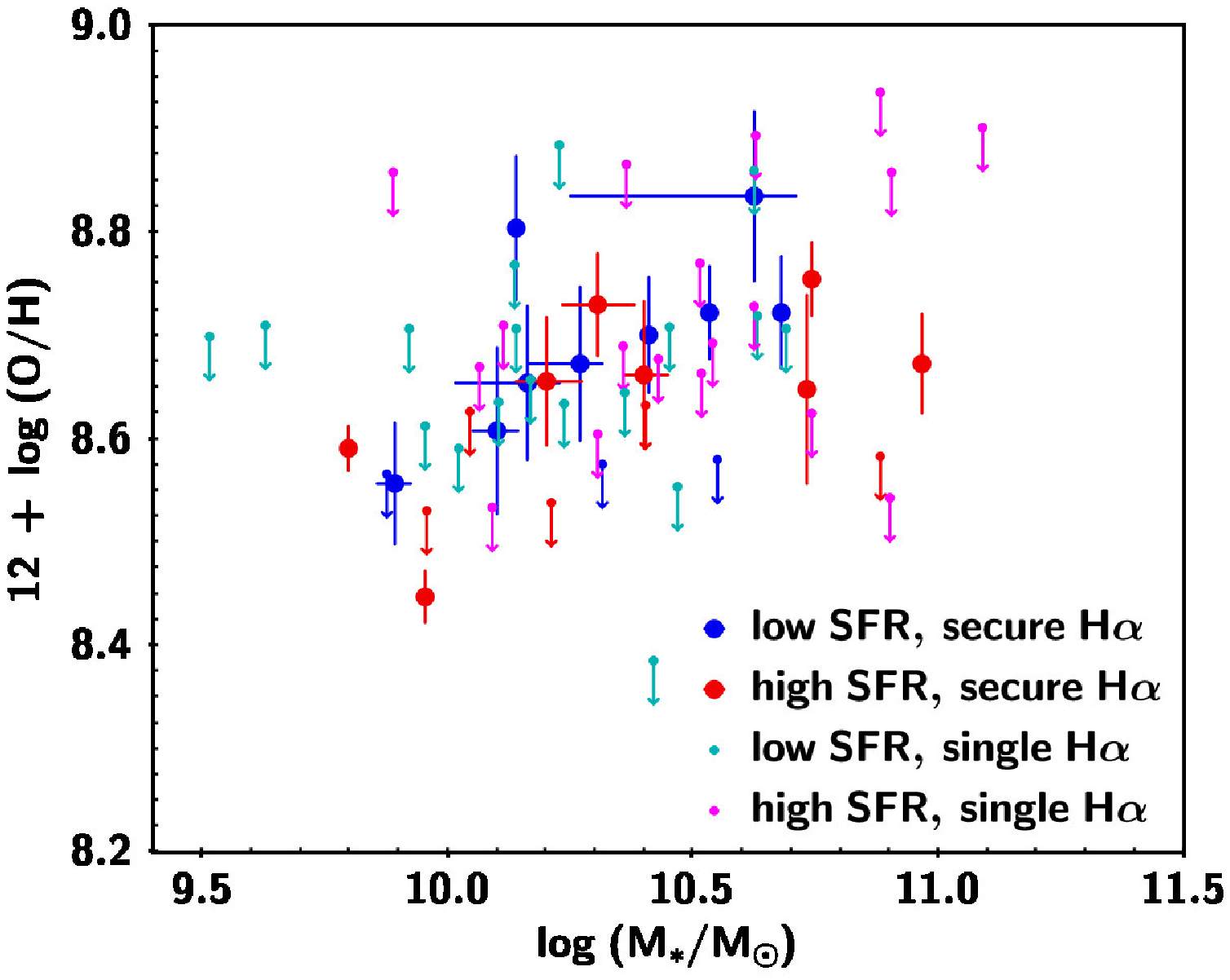}
    \caption{The dependence of the mass--metallicity relation on the SFR. Galaxies in our sample are divided into two subsamples of high- and low-SFR shown in redder and bluer color-coded indices, respectively. The bigger and smaller indices represent the secure and single H$\alpha$ detected sources, respectively.}
    \label{fig:MZ-SFRdependent}
  \end{center}
\end{figure}
We examine the dependence of the mass--metallicity relation on SFR for our sample at $z\sim0.88$.
In Figure \ref{fig:MZ-SFRdependent}, we present the metallicity as a function of the stellar mass, dividing two subsamples according to their SFR.
The median SFRs of the high-SFR and low-SFR subsamples are 35.0~M$_{\Sol}$~yr$^{-1}$ and 16.8~M$_{\Sol}$~yr$^{-1}$, respectively.
In Figure \ref{fig:MZ-SFRdependent}, we do not see evidence of SFR dependence of the mass--metallicity relation for the $z\sim 0.88$ IR galaxies.

Our sample only covers the SFR range of 0.8$<$~log(SFR/M$_{\Sol}{\rm yr}^{-1}$)~$<$~1.9, while the sample at $z\sim0.1$ by \citet{Mannucci10} covers the SFR range of $-$1.45~$<$~log(SFR/M$_{\Sol}{\rm yr}^{-1}$)~$<$~0.8. 
Thus the SFR of our sample is larger and there is almost no overlap of SFRs with the local sample.
According to Figure 1 of \citet{Mannucci10}, there is a tendency that the higher stellar mass is, the less remarkable the dependence on SFR is.
At $M_{\ast}=10^{10}~{\rm M}_{\Sol}$, a galaxy typically has metallicity 0.1~dex smaller than that of one with one order of magnitude larger SFR.
Because of the SFR range of our sample, an expected metallicity difference between galaxies with the maximum and the minimum SFRs is about 0.1~dex, which is comparable to the metallicity dispersion of our sample (see \S \ref{sec:MZrelation}).
Therefore, even if there is a SFR dependence of the mass--metallicity relation, it is probably not detectable with our sample.
A sample covering a larger SFR range is required to investigate the dependence for large stellar mass galaxies.

\subsection{The fundamental metallicity relation}
\label{sec:FMR}

\begin{figure}[htbp]
  \begin{center}
    \includegraphics[width=55mm, angle=-90]{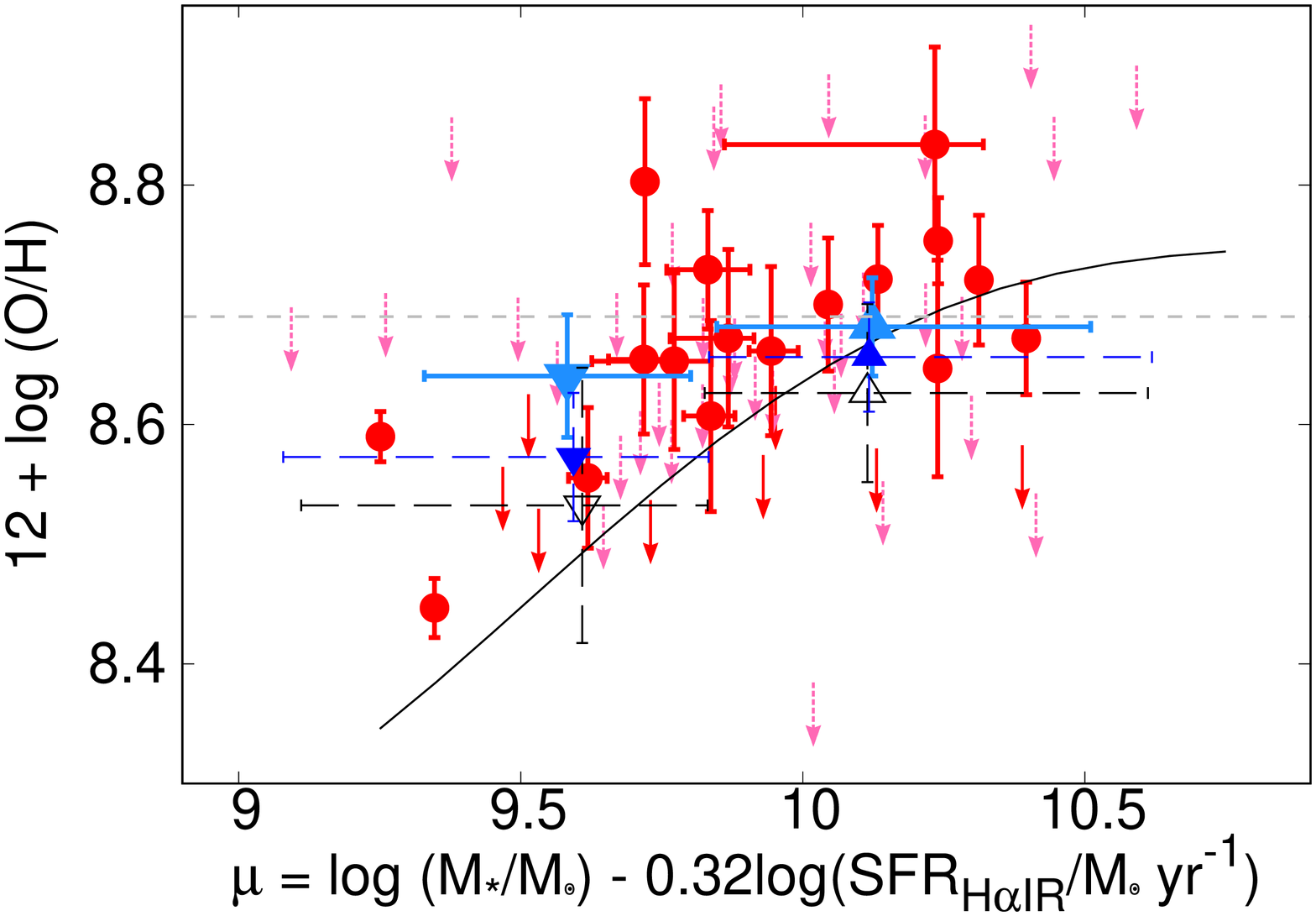}
    \includegraphics[width=55mm, angle=-90]{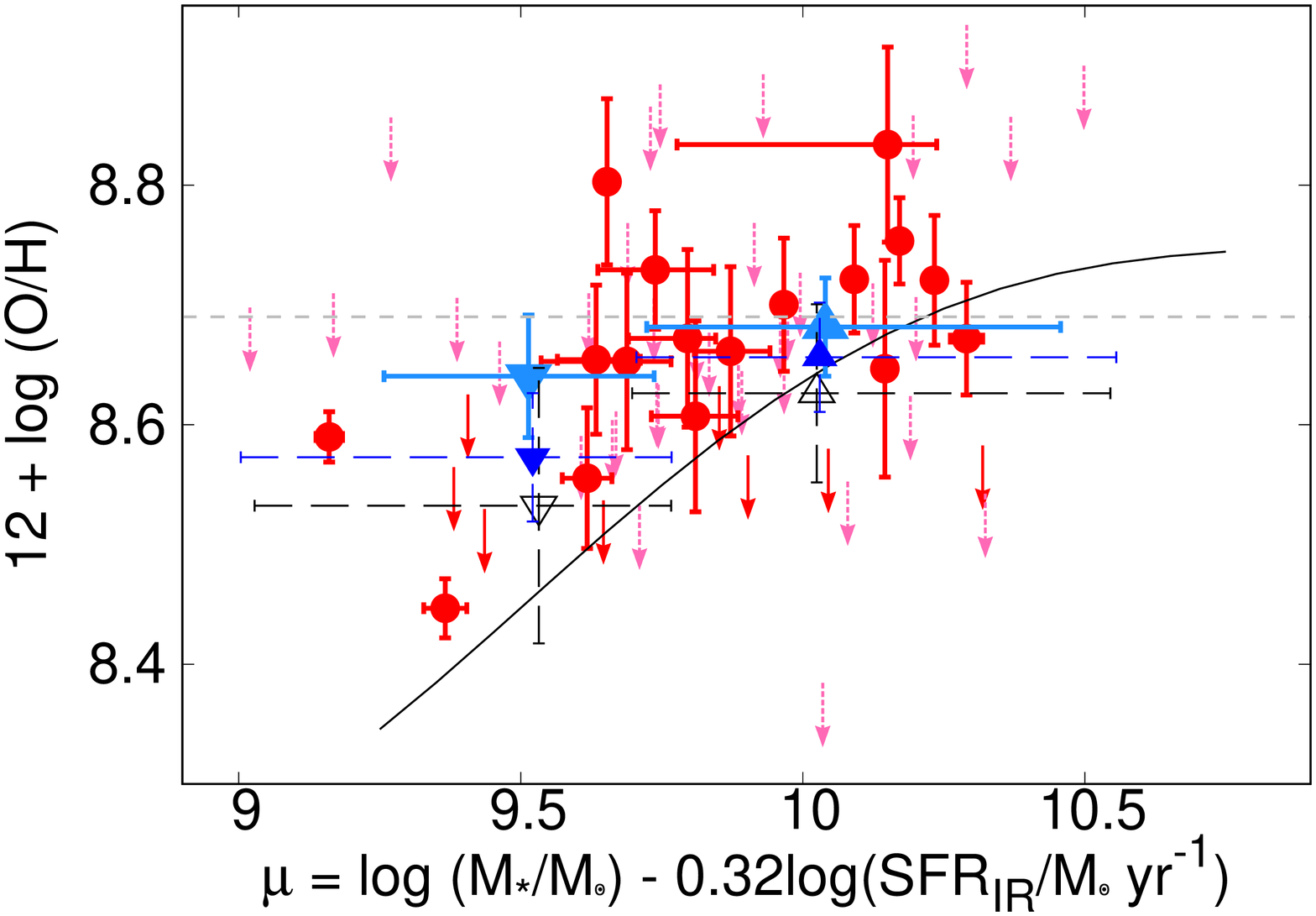}
    \includegraphics[width=55mm, angle=-90]{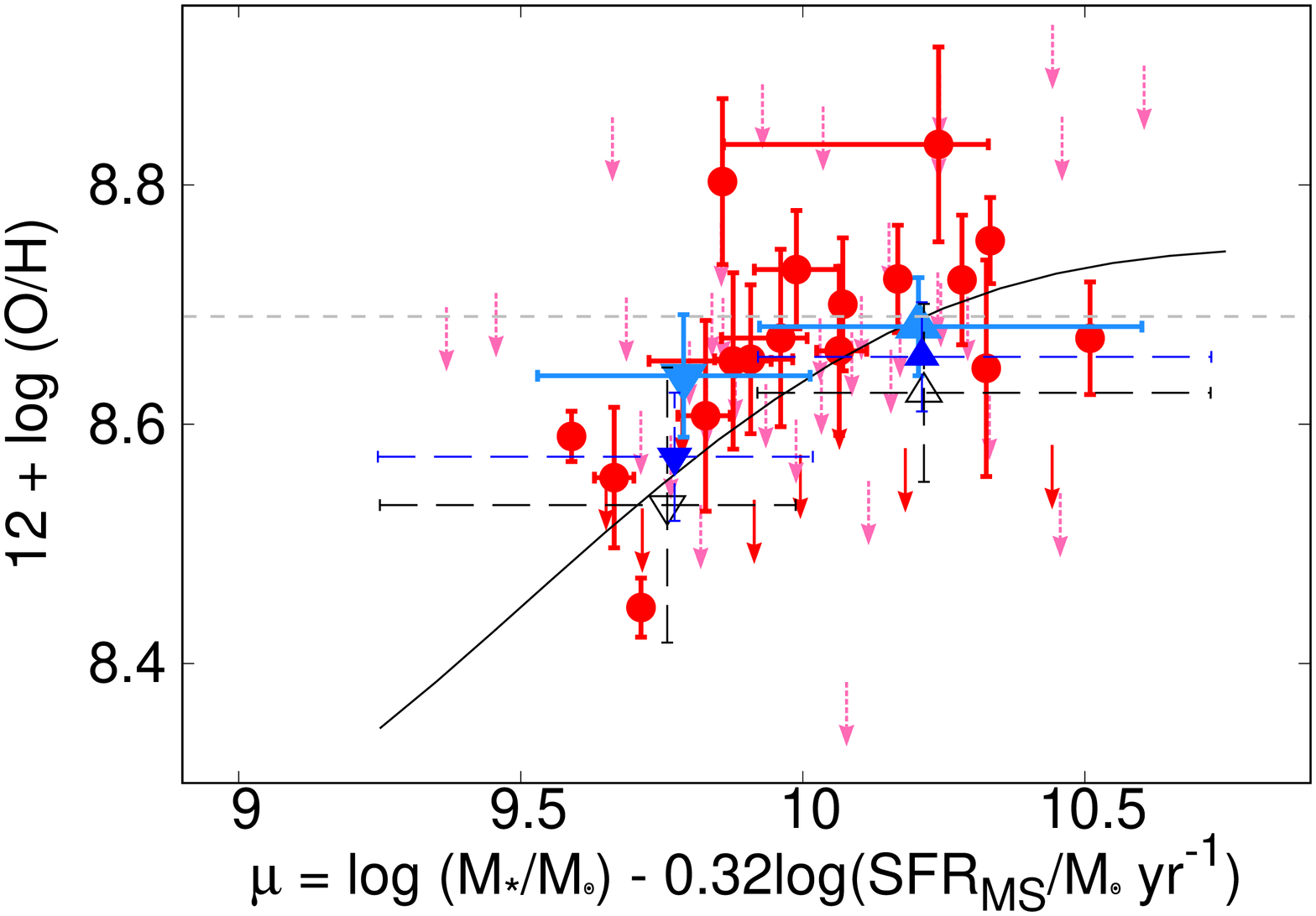}
    \caption{ Metallicity distribution against the projection axis $\mu$ of the FMR with a projection parameter $\alpha$ of 0.32. The symbols of our sample are the same as Figure \ref{fig:MZrelation}. The projection axis $\mu$ for each galaxy is calculated with its SFR. The FMR by \citet{Mannucci10} is shown by the black solid line with conversion to N2 method.
 (Right) The same as the left panel, but SFRs are calculated using only infrared (without using H$\alpha$). 
 (Bottom) The same as the left panel, but SFRs of MS galaxies at $z\sim 0.78$ \citep{Zahid12} at the stellar mass of our sample are used to measure the projection axis $\mu$.}
    \label{fig:FMZ}
  \end{center}
\end{figure}

In the left panel of Figure \ref{fig:FMZ}, we plot the metallicity of our IR galaxies and stacking analysis results against a projected axis $\mu~\equiv$ log($M_{\ast}$)$-\alpha$~log(SFR), where $\alpha$ is a projection parameter on the plane.
We adopt $\alpha$ of 0.32 which minimizes the intrinsic scatter of metallicity for local SDSS galaxies \citep{Mannucci10}.
For comparison, we convert the FMR whose metallicity is based on \citet{Nagao06} and \citet{Maiolino08} to the N2 method (solid curve in Figure \ref{fig:FMZ}).
We find that the metallicity of stacking for massive galaxies (blue upward triangle) comes on the FMR within the error bars, whereas that for less massive galaxies (blue downward triangle) shows a 2$\sigma$ offset from the FMR at a fixed $\mu$. 
The offset is not due to a particular SFR derivation as the right panel, where SFR is computed only using infrared, shows a similar offset (right panel of Figure \ref{fig:FMZ}).
  
We have found in Figure \ref{fig:MS} that our sample at $z\sim0.88$ has an average SFR of 25~M$_{\Sol}~{\rm yr}^{-1}$, 
which is significantly above that of MS galaxies at $z\sim0.78$ at a given mass.
By contrast, we have found in Figure \ref{fig:MZrelation} that the mass--metallicity relation of our sample agrees with that of the $z\sim0.78$ MS galaxies within the error bars.
Therefore we suspect that the reason of the big difference of our sample from the FMR in Figure \ref{fig:FMZ} is the difference of SFR.  
To test that, we again plot our galaxies in the 3-D space with SFRs which are predicted for the MS galaxies at $z\sim0.78$ for
the stellar masses of our galaxies (bottom panel of Figure \ref{fig:FMZ}). 
This shows that our galaxies having similar SFRs to MS galaxies matches well with the local FMR.
The discrepancy between our sample and the FMR is caused by stronger star-formation. 
Thus, we conclude that the IR galaxies at $z\sim 0.88$ in our sample have similar metallicity but star-formation is elevated 
at the time of observation, compared with typical star-forming galaxies.  

If the discrepancy between our results and the FMR is real, then a possible interpretation is seeing galaxies in a merging phase.
\citet{Cox08} has simulated a merger-driven starburst and found that SFR of a merging galaxy of $M_{\ast}>10^{10}~{\rm M}_{\Sol}$ is dramatically and instantaneously enhanced by ten times or more, and reaches to $\sim$20~${\rm M}_{\Sol}~{\rm yr}^{-1}$, which is consistent with our sample. 

According to the suggestion of \citet{Dave11}, however, a galaxy merger not only increases the SFR but also decreases the metallicity because of fresh gas inflow, since smaller galaxies have lower metallicity.
Indeed, some observations have found lower metallicities in merger systems.  
\citet{Ellison08b} found that closer pair galaxies in the local Universe have lower metallicities.
\citet{Peeples09} found that galaxies with significantly lower metallicity than the mean mass--metallicity relation at $z\sim0$ tend to be showing signs of galaxy interactions.
Our IR galaxies have higher SFR, but not lower metallicity.
Taking account of the slower metallicity enrichment compared to the timescale of metal dilution, our result may imply that gas falling into the IR galaxies in our sample may not be metal-poor.

Environments may seem like a natural explanation for the lack of metallicity dilution of our sample.
Galaxy evolution is supposed to progress faster in higher density regions.
Cooper et al. (2008a, b) found that more metal-rich galaxies tend to be in higher-density regions such as centers of groups or clusters of galaxies.
If our galaxies are located in such high-density regions, it is possible that the environments lessen metal dilutions by gas infall from the IGM.
Unfortunately the lack of redshift information for galaxies in the environs of our sample precludes examining the environment effects.

Another possible explanation is a major merger of approximately equal-mass galaxies. 
Their metallicities are expected to be similar to each other, preventing metallicity dilution by gas infall.
SFR enhancement can be higher for merger galaxies of massive and similar stellar mass \citep{Cox08}.
Since our FMOS detection limits us toward higher SFR, we might preferentially observe galaxies 
in some stage of merger between roughly equal mass galaxies.

\subsection{Comparison with other IR galaxy samples}
There are some previous investigations of mass--metallicity relations for IR galaxies.
\citet{Rupke08} studied metallicity abundance of 100 LIRGs and ULIRGs at $z\sim0.1$ which were basically selected from IRAS Survey catalog, and found that nuclei of the IR galaxies are under-abundant by a factor of 2 compared with normal star-forming galaxies at the same luminosity and stellar mass.
\citet{Kilerci-Eser14} studied 118 ULIRGs and HLIRGs at $z\sim0.18$ from the AKARI All-Sky survey catalog, and confirmed the lower metallicity tendency. 
Since these local IR galaxies show interaction features such as multiple galaxy systems, tidal tails or bridges, and disturbed morphology, the sources are considered as ongoing or pre-mergers. 
On the other hand, \citet{Roseboom12} investigated 57 $Herschel$/SPIRE sources at $z\sim1.2$ and concluded that the mass--metallicity relation of their sample is indistinguishable from the local mass--metallicity relation, which is consistent with the results of our IR galaxies at $z\sim0.88$.
Results of these studies suggest that there is a discrepancy of behavior of the mass--metallicity relations between IR galaxies at low and high redshift compared with MS galaxies.
We note that SFRs of these IR galaxies are higher than MS galaxies at similar redshifts of each IR galaxy sample.
It has been known that interactions and mergers are much more common among IR galaxies (e.g., \cite{Veilleux02}).
If the IR galaxies at high-$z$ Universe are also in some merger phase, the inflowing gas may not be as pristine as for local IR galaxies. 

\section{Summary}
We have investigated the relation among stellar mass, gas-phase oxygen abundance, and SFR using Subaru/FMOS spectroscopy of AKARI-detected mid-IR galaxies at $\langle z\rangle \sim0.88$ in the NEP.
We have observed $\sim$350 AKARI sources and have 28 secure H$\alpha$ emission detections and 44 possible H$\alpha$ emission detections. Three sources out of the 28 secure H$\alpha$ detected sources are affected by an AGN, so we excluded the three objects from the following discussions.

We have measured stellar mass using SED fitting to photometric data from optical $u^{\ast}$-band to far-IR 500$\mu$m band, SFRs using $L_{\rm H\alpha}$ and $L_{\rm IR}$, and metallicity from the [N \emissiontype{II}]/H$\alpha$ emission line ratio.
The following are the main results and conclusions of our analysis.

\begin{enumerate}
\item SFRs of our sample are almost constant over the stellar mass range of our sample ($10^{9.52}-10^{11.09}~{\rm M}_{\Sol}$) due to the detection limit of our observation.
The average SFR is 25~${\rm M}_{\Sol}~{\rm yr}^{-1}$ which is higher for less massive galaxies ($\sim$10$^{10.05}~{\rm M}_{\Sol}$) by $\sim$0.6~dex than that of MS galaxies at $z\sim0.78$, while which is consistent for massive galaxies ($\sim$10$^{10.58}~{\rm M}_{\Sol}$) 
compared to that of the MS galaxies.

\item A positive correlation between stellar mass and metallicity of our IR galaxies is found, consistent with previous studies. 
The derived best-fit mass--metallicity relation for metallicity determined galaxies agrees with the relation of optically selected normal star-forming galaxies in the local Universe.
On the other hand, the metallicities measured from stacked spectra in two stellar mass bins agree with those at similar stellar masses in a similar redshift to that of ours. 
The results suggest that typical metallicity of IR galaxies in our sample is not significantly different from the MS galaxies at similar redshift, and metal-rich galaxies are already metal-enriched to the level of local galaxies up to then. 

\item No discernible dependence of the mass--metallicity relation on SFR is seen.
The SFR range in our sample would only predict a metallicity difference of $\sim$0.1~dex, comparable to the scatter of our mass--metallicity relation.
Thus, it is difficult to discuss the SFR dependency with our sample.

\item The metallicity of our sample as a function of projection axis $\mu$ of ${\rm log}(M_{\ast})-0.32{\rm log}({\rm SFR})$ shows a significant offset from the FMR suggested by \citet{Mannucci10}, especially for less massive galaxies.
Whereas, when we adopt SFRs of MS galaxies at stellar masses of our sample to measure the prediction axis, the metallicity distribution shows agreement with the FMR.
Since our sample galaxies show higher SFRs but no under-abundance, we consider
that this may be caused by mergers.
\end{enumerate}

\begin{ack}
We would like to thank an anonymous referee for useful comments. 
This work is supported by the Japan Society for the Promotion of Science (JSPS; grant number 23244040).
TG acknowledges the support by the Ministry of Science and Technology of Taiwan through grant NSC 103-2112-M-007-002-MY3, and
105-2112-M-007-003-MY3.
We are grateful to the FMOS support astronomers I. Tanaka and K. Aoki for their support during the observations.
We thank F. Iwamuro, K. Yabe, K. murata, B. Vuat for helpful discussions. 
We would like to express our acknowledgement to the indigenous Hawaiian people for their understanding of the significant role of the summit of Maunakea in astronomical research.
\end{ack}

\appendix

\begin{table}[p]
\begin{center}
\caption{Information of no secure-H$\alpha$ detected sources.\hfil\hfill}
\label{tb:single}
\begin{tabular}{cccccc}
\hline
  \multicolumn{1}{c}{Object} &
  \multicolumn{1}{c}{RA} &
  \multicolumn{1}{c}{DEC} &
  \multicolumn{1}{c}{Wavelength}&
  \multicolumn{1}{c}{Redshift} &
  \multicolumn{1}{c}{Flux} \\
  \multicolumn{1}{c}{} &
  \multicolumn{1}{c}{[deg]} &
  \multicolumn{1}{c}{[deg]} &
  \multicolumn{1}{c}{[$\mu$m]} &
  \multicolumn{1}{c}{} &
  \multicolumn{1}{c}{[$10^{-16}$~erg/s/cm$^{2}$]} \\
\hline
  \multicolumn{6}{c}{Single emission line detected sources} \\
\hline
  61003682 & 268.87315 & 66.18124 &           1.3333 &            1.032 &  1.11$\pm$0.15  \\
  61004486 & 269.49407 & 66.22477 &           1.1226 &            0.710 &  0.80$\pm$0.15 \\
  61005062 & 269.13717 & 66.24974 &           1.2751 &            0.943 &  0.79$\pm$0.58 \\
  61005551 & 269.16760 & 66.27146 &           1.3346 &            1.034 &  0.90$\pm$0.14 \\
  61006497 & 268.51170 & 66.30662 &           1.2913 &            0.968 &  0.41$\pm$3.44 \\
  61007099 & 268.29289 & 66.33036 &           1.2615 &            0.922 &  1.84$\pm$0.27 \\
  61007665 & 269.34977 & 66.35227 & $\sim$1.2141 & $\sim$0.850 & $\cdots$             \\
  61007744 & 268.92021 & 66.35504 &           1.2609 &            0.921 &  0.64$\pm$0.20 \\
  61008501 & 268.83375 & 66.37901 &           1.3346 &            1.034 &  0.28$\pm$0.09 \\
  61009352 & 268.98370 & 66.40349 &           1.2484 &            0.902 &  0.82$\pm$0.14 \\
  61009999 & 269.05553 & 66.42234 &           1.2742 &            0.941 &  1.65$\pm$0.27 \\
  61010363 & 268.77946 & 66.43416 &           1.2481 &            0.902 &  0.58$\pm$0.12 \\
  61011753 & 269.23698 & 66.47298 &           1.2607 &            0.921 &  0.61$\pm$0.15 \\
  61012181 & 268.87003 & 66.48556 & $\sim$1.2241 & $\sim$0.865 & $\cdots$            \\
  61012206 & 268.34379 & 66.48484 &           1.3299 &            1.026 &  0.63$\pm$0.22 \\
  61013146 & 269.26709 & 66.51236 &           1.1863 &            0.808 &  0.59$\pm$0.21 \\
  61013508 & 268.15150 & 66.51896 & $\sim$1.3088 & $\sim$0.994 &  $\cdots$            \\
  61014814 & 269.13637 & 66.55578 &           1.3322 &            1.030 &  0.22$\pm$0.11 \\
  61014853 & 269.09904 & 66.55698 &           1.1279 &            0.719 &  0.73$\pm$0.69 \\
  61015291 & 268.87158 & 66.56778 &           1.3134 &            1.001 &  0.99$\pm$0.13 \\
  61015448 & 268.91175 & 66.57122 &           1.3144 &            1.003 &  1.02$\pm$0.23 \\
  61015541 & 268.72560 & 66.57175 &     $<$1.1250 &      $<$0.714 &  $\cdots$          \\
  61016430 & 269.00271 & 66.59490 &           1.3148 &           1.003 & 1.45$\pm$0.27  \\
  61016436 & 268.74498 & 66.59369 &           1.2034 &           0.834& 0.22$\pm$0.22  \\
  61016569 & 268.43110 & 66.59759 & $\sim$1.3073 & $\sim$0.992& $\cdots$            \\
  61016583 & 269.07232 & 66.59913 &           1.1796 &           0.797& 1.94$\pm$0.31  \\
  61017368 & 268.75418 & 66.62051 &           1.2780 &           0.947& 0.32$\pm$0.06  \\
\hline
\end{tabular}
\end{center}
Information of no secure-H$\alpha$ detected sources. In a case that a peak of a emission line could not be measured due to OH masks or out of observation wavelength range, approximate value is listed. Redshift is estimated with an assumption that the detected line is H$\alpha$.
\end{table}

\begin{table}[p]
\begin{center}
\caption{Information of no secure-H$\alpha$ detected sources.\hfil\hfill}
\begin{tabular}{cccccc}
\hline
  \multicolumn{1}{c}{Object} &
  \multicolumn{1}{c}{RA} &
  \multicolumn{1}{c}{DEC} &
  \multicolumn{1}{c}{Wavelength}&
  \multicolumn{1}{c}{Redshift} &
  \multicolumn{1}{c}{Flux} \\
  \multicolumn{1}{c}{} &
  \multicolumn{1}{c}{[deg]} &
  \multicolumn{1}{c}{[deg]} &
  \multicolumn{1}{c}{[$\mu$m]} &
  \multicolumn{1}{c}{} &
  \multicolumn{1}{c}{[$10^{-16}$~erg/s/cm$^{2}$]} \\
\hline
  \multicolumn{6}{c}{Single emission line detected sources} \\
\hline
  61017497 & 268.20317 & 66.62255 &           1.3286 &           1.024& 1.32$\pm$0.23  \\
  61017713 & 269.15508 & 66.63112 &           1.3273 &           1.022& 0.60$\pm$0.18  \\   
  61017884 & 268.61201 & 66.63417 & $\sim$1.1993 & $\sim$0.827&  $\cdots$           \\   
  61018263 & 268.70380 & 66.64585 &           1.3067 &            0.991& 0.66$\pm$0.11 \\  
  61018472 & 269.34189 & 66.65131 &           1.3268 &            1.022& 0.79$\pm$0.14 \\
  61019363 & 268.73456 & 66.67666 &           1.2627 &            0.924& 1.88$\pm$0.25 \\
  61020455 & 268.99704 & 66.70688 &           1.2010 &            0.830& 0.82$\pm$0.13 \\
  61021137 & 269.24127 & 66.72676 &           1.2793 &            0.949& 0.34$\pm$0.13 \\   
  61022730 & 268.89905 & 66.77878 &           1.2839 &            0.956& 0.71$\pm$0.21 \\
  61023133 & 268.64746 & 66.79599 &           1.1251 &            0.714& 0.55$\pm$0.12 \\
  61023218 & 269.35485 & 66.80129 & $\sim$1.3116 & $\sim$0.998& $\cdots$           \\
  61023314 & 268.83595 & 66.80321 &           1.2636 &            0.925& 0.78$\pm$0.23\\
  61023461 & 269.13738 & 66.81022 &           1.2031 &            0.833& 2.54$\pm$0.22 \\
  61023651 & 268.71468 & 66.81580 &           1.1253 &            0.715& 0.68$\pm$0.09\\
  61024161 & 268.78796 & 66.83644 &           1.2639 &            0.926& 0.88$\pm$0.15\\
  61024948 & 269.06189 & 66.87105 &           1.2839 &            0.956& 0.82$\pm$0.18\\
  61025402 & 268.79856 & 66.89269 & $\sim$1.3121 &            0.999&  $\cdots$        \\
\hline
  \multicolumn{6}{c}{[O \emissiontype{III}] lines detected source} \\
\hline
  61012430 & 269.01647 & 66.49320 & 1.2244, 1.2364 &         1.4693 & 1.63$\pm$0.27, 0.64$\pm$0.18\\
\hline
  \multicolumn{6}{c}{[O \emissiontype{II}] lines detected source} \\
\hline
  61020689 & 269.38283 & 66.71269 & 1.2793, 1.2804 &        2.5282 & 0.51$\pm$0.11, 0.52$\pm$0.06\\
  61022567 & 269.10734 & 66.77444 & 1.1771, 1.1781 &       2.1592 & 0.44$\pm$0.07, 0.32$\pm$0.06\\
\hline
\end{tabular}
\end{center}
Continue of Table \ref{tb:single}, There is one source whose emission lines are [O \emissiontype{III}] doublet, and two sources whose emission lines are [O \emissiontype{II}]doublet. In the case, the redshifts are calculated with the rest-frame wavelength of these emission lines.
\end{table}

\end{document}